\documentclass[aip,jap,reprint,superscriptaddress,showpacs,letter]{revtex4-1}

\usepackage{graphicx,psfrag}
\usepackage[latin1]{inputenc}
\usepackage{natbib}
\usepackage{amsmath, amsthm, amssymb}
\usepackage{color}
\usepackage{comment}
\usepackage{ulem}

\newcommand{\notes}[1]{}

\newcommand{\beq}{\begin{equation}}
\newcommand{\eeq}{\end{equation}}
\newcommand{\beqnn}{\begin{equation*}}
\newcommand{\eeqnn}{\end{equation*}}
\newcommand{\beqas}{\begin{eqnarray*}}
\newcommand{\eeqas}{\end{eqnarray*}}
\newcommand{\beqa}{\begin{eqnarray}}
\newcommand{\eeqa}{\end{eqnarray}}

\begin{document}

\title{Temperature-Dependent Anisotropic Magnetoresistance and Spin-Torque-Driven Vortex Dynamics in a Single Microdisk}
\date{\today}
\author{Sergi Lend\'inez}
\affiliation{Materials Science Division, Argonne National Laboratory, Argonne, Illinois 60439, USA}
\affiliation{Department of Physics and Astronomy, University of Delaware, Newark, DE 19716, USA}

\author{Tomas Polakovic}
\affiliation{Physics Division, Argonne National Laboratory, Argonne, Illinois 60439, USA}
\author{Junjia Ding}
\affiliation{Materials Science Division, Argonne National Laboratory, Argonne, Illinois 60439, USA}
\author{M. Benjamin Jungfleisch}
\email{mbj@udel.edu}
\affiliation{Department of Physics and Astronomy, University of Delaware, Newark, DE 19716, USA}
\author{John Pearson}
\affiliation{Materials Science Division, Argonne National Laboratory, Argonne, Illinois 60439, USA}
\author{Axel Hoffmann}
\altaffiliation[Current address: ]{Department of Materials Science and Engineering, University of Illinois at Urbana-Champaign, Urbana, IL 61801, USA}
\affiliation{Materials Science Division, Argonne National Laboratory, Argonne, Illinois 60439, USA}
\author{Valentine Novosad}
\affiliation{Materials Science Division, Argonne National Laboratory, Argonne, Illinois 60439, USA}
\email{novosad@anl.gov}

\begin{abstract}
Spin-orbit-torque-driven dynamics have recently gained interest in the field of magnetism due to the reduced requirement of current densities and an increase in efficiency, as well as the ease of implementation of different devices and materials. 
From a practical point of view, the low-frequency dynamics below 1 GHz is particularly interesting since dynamics associated with magnetic domains lie in this frequency range. While spin-torque excitation of high-frequency modes has been extensively studied, the intermediate low-frequency dynamics have received less attention, although spin torques could potentially be used for both manipulation of the spin texture, as well as the excitation of dynamics. In this work, we demonstrate that it is possible to drive magnetic vortex dynamics in a single microdisk by spin-Hall torque at varying temperatures, and relate the results to transport properties. We find that the gyrotropic mode of the core couples to the low-frequency microwave signal and produces a measurable voltage. The dynamic measurements are in agreement with magnetic transport measurements and are supported by micromagnetic simulations. 
Our results open the door for integrating magnetic vortex devices in spintronic applications.
\end{abstract}

\maketitle

\section{Introduction}\label{sec:introduction}

Nowadays, spin-orbit torque is a commonly used phenomenon in magnetic materials to control the magnetization state. It has been used to switch the magnetization\cite{Liu2012, Liu2012a, Fukami2016}, excite  ferromagnetic resonance \cite{Liu2011}, move domain walls in racetrack memories \cite{Parkin2008, Parkin2015}, and to produce auto-oscillations in magnetic heterostructures \cite{Liu2013, Demidov2014, Demidov2014a}, which has been discussed as potential mechanism for neuromorphic computing applications \cite{Locatelli2014}. Spin torques offer a more efficient way to control the magnetization than other methods such as Oersted fields, in which relatively high currents are needed. As opposed to the use of Oersted fields, spin-torques allow the switching and control of the magnetization in a much more localized region, which allows higher density devices \cite{Brataas2012}. Moreover, spin-orbit torques enable a simpler fabrication and detection \textcolor{black}{than in typical multi-layer spin-torque devices}, as bilayer stacks can be utilized and magnetization dynamics in low-damping magnetic insulators can be excited \textcolor{black}{in a lateral geometry}\cite{Baumgartner2017,Manchon2019}.

In some magnetic systems, the onset of auto-oscillations  can be achieved when the spin-orbit-torque effect is large enough to compensate damping in a magnetic material.  
It was shown that this can be realized in ferromagnet/heavy metal bilayers with a constricted area, where the current density is locally enhanced\textcolor{black}{\cite{Demidov2014,Demidov2014a}}. A spin current that is created by means of the spin Hall effect in the heavy metal interacts with the magnetization in the ferromagnet by exerting a torque and driving the system into auto-oscillations. Moreover, several auto-oscillators can be coupled together, producing an even higher output microwave power \cite{Awad2016}. 

Most of the spin-orbit-torque studies so far have focused either on the control of the magnetization (including the movement of magnetic textures such as skyrmions \cite{Woo2017,Jiang2017}) or on the excitation of high-frequency dynamics in the GHz range. However, the low-frequency dynamics of several hundreds of MHz up to one GHz has not been explored by spin-orbit torques in such detail. This intermediate low-frequency response is particularly interesting since magnetic domains and magnetization textures oscillate in this frequency range. For instance, dynamics associated with the motion of magnetic vortices lie in this sub-one GHz range and have been studied in detail by microwave and transport measurements \cite{Cowburn1999, Choe2004, Buchanan2005, Kasai2006, Caputo2007, Sugimoto2011, Pollard2012}. One of its important properties is that the individual ground state is stable even at high temperatures, and it can be controlled by applying an external magnetic field pulse \cite{VanWaeyenberge2006}. Recently, it was shown that the dynamics in an array of magnetic disks can be used for spin-pumping applications \cite{Hasegawa2017}. Furthermore, it was shown that a spin-ice network made of interacting magnetic microdisks has a controllable ground state, depending on the excitation frequency \cite{Behncke2018}. 

Here, we demonstrate that spin torque can even drive low-frequency dynamics in a single disk made of a Ni$_{80}$Fe$_{20}$/Pt bilayer, and that the oscillation couples to the time-varying anisotropic magnetoresistance of the disk, leading to a measurable voltage. This mechanism is in analogy to spin-torque ferromagnetic resonance, but in the sub-one GHz regime. Furthermore, we carry out detailed micromagnetic simulations and correlate our experimental findings with magnetic transport measurements. 
Our results are the first steps towards the excitation of auto-oscillations of the magnetic vortex in these kinds of structures.

This article is structured in the following way. In the first part (Sec.~\ref{sec:dc}), we characterize the behavior of a single microdisk under the effect of an external magnetic field at temperatures from 2~K to 300~K. Previous studies have shown that the nucleation and annihilation of a magnetic vortex \textcolor{black}{as a function of an external magnetic field could be characterized measuring the anisotropic magnetoresistance (AMR) of the disk with a dc electrical current (dc-AMR)\cite{Kasai2006}, or even by measuring the voltage in a cross-bar Hall sensor produced by a nanodisk deposited on top of the sensor \cite{Mihajlovic2010}. In our work, we characterize the disk using the dc-AMR trace. 
}
 Hence, in order to improve the \textcolor{black}{dc-}AMR signal in our experiment, we fabricated a 50~nm~thick disk consisting of Ni$_{80}$Fe$_{20}$ only. The details of the fabrication process as well as the experimental setup are given in section \ref{sec:dcsetup}, and the results of the dc measurements in section \ref{sec:dcresults}.

The second part of this work (Sec. \ref{sec:rf}) discusses the dynamics of a single microdisk using a rf homodyne detection technique based on the spin-torque ferromagnetic resonance (ST-FMR) \cite{Liu2011}, in which a Pt layer is used to create a spin current. For this purpose, a 5-nm-thick Pt layer is fabricated on top of a 35-nm-thick Ni$_{80}$Fe$_{20}$ disk. The lower Ni$_{80}$Fe$_{20}$ thickness compared to the first part of the paper is chosen to enhance the  
signal as discussed below. The details of the fabrication and the setup for the rf measurements are given in section \ref{sec:rfsetup}, and the corresponding results are presented in section \ref{sec:rfresults}.

Finally, in section \ref{sec:conclusion} we discuss the conclusions of our study.

\section{Vortex Formation}\label{sec:dc}

In order to characterize the magnetic behavior of a single microdisk, we use dc transport measurements to experimentally obtain the AMR. We then compare the experimental data to the results of micromagnetic simulations to deduce the magnetic configuration. In order to improve the signal strength, we choose a thickness of 50 nm for the magnetic material with no heavy metal capping layer.

\subsection{Experimental setup for dc measurements}\label{sec:dcsetup}

\begin{figure}
\includegraphics[width=\columnwidth]{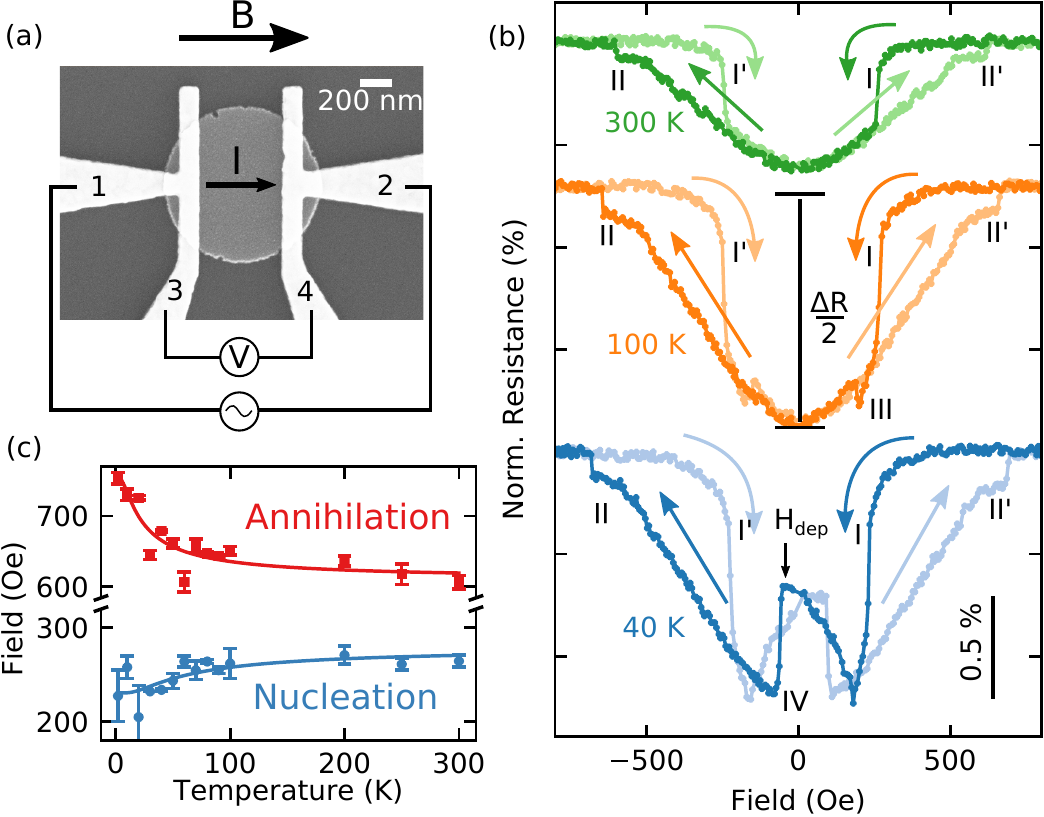}
\caption{(a) SEM image of a 1~$\mu$m-diameter Py disk and Ti/Au nanocontacts to measure magnetotransport. The current was applied through the external contacts (1 and 2), and the voltage response is measured on the middle contacts (3 and 4) with a lock-in technique. Magnetic field was applied parallel to the current. (b) AMR curves of the disk measured with a 1.4 kHz sinusoidal current with an amplitude of \textcolor{black}{100}~$\mu$A at 300~K (top, green), 100~K (middle, orange), and 40~K (bottom, blue). Dark colors show the down field sweep, and light colors show the up field sweep, as indicated by the arrows. The roman numbers indicate the important features of the curves: as field is swept down a vortex is nucleated (I), annihilated (II), and intermediate states appear (III and IV). Equivalent features are observed when the field is swept up (I' and II') (c) Temperature dependance of the nucleation (blue) and annihilation (red) fields, extracted from the AMR curves as shown in panel (b). As temperature increases, the nucleation field monotonically decreases, whereas the nucleation field monotonically increases.\label{fig:dcexp}}
\end{figure}

Our samples were fabricated using a multistep electron-beam lithography (EBL) process. First, the disk with 1 $\mu$m diameter and alignment marks were defined on a positive bilayer resist of ZEP520A and PMGI SF2 on a silicon substrate, accompanied by e-beam evaporation and lift-off process. 50 nm of Ni$_{80}$Fe$_{20}$ (Permalloy, Py) was deposited on the samples for the temperature-dependent AMR study. The second step consists of patterning of contacts to the sample, followed by deposition of 5 nm Ti and 100 nm Au and lift-off, completing the fabrication process. The scanning electron microscopy (SEM) image [Fig. \ref{fig:dcexp}(a)] shows a good alignment between the different steps, as well as a high quality of the Au contacts and the Py microdisk. 

The dc electrical response of the disk was measured with 4-probe electrical contacts, marked as 1 to 4 in Fig. \ref{fig:dcexp}(a). The minimum width of the contact electrodes -- corresponding to 3 and 4 in Fig.~\ref{fig:dcexp}(a)  -- is 150 nm, while the distance between those contacts is 500 nm. The contacts extend along the entire disk so that a more uniform current distribution along the disk is obtained. The contacts connect to square pads of 100 $\mu$m$^2$, to which the sample is wire-bonded using Al bonds. The samples were then introduced in a physical property measurement system (PPMS) cryostat capable of reaching temperatures down to 2~K and magnetic fields up to 5~T. A 1.4 kHz sine-modulated current of \textcolor{black}{100} $\mu$A in amplitude was applied to the disk through ports 1 and 2, and the voltage response was picked up by a lock-in amplifier through ports 3 and 4. The signal gain was 500, achieved by a low-noise pre-amplifier. Our disks had a typical resistance value \textcolor{black}{
around 3 $\Omega$}, which corresponds to a resistivity value of $\rho\sim50$ $\mu\Omega\cdot$cm, in good agreement with the known Py resistivity\cite{Mayadas1974}. The resistance was measured as a function of an in-plane magnetic field parallel to the current in the range from $-1000$~Oe to $+1000$~Oe for different temperatures from 2~K to 300~K.

\subsection{Results of dc measurements}\label{sec:dcresults}
The change in resistance is produced by the AMR of the device. AMR is the dependence of the resistivity on the relative orientation between the magnetization and the electric current:
\begin{equation}\label{amr}
\rho=\rho_\perp+(\rho_\|-\rho_\perp)(\mathbf{j}\cdot\mathbf{m})^2=\rho_\perp+(\rho_\|-\rho_\perp)\mathrm{cos}^2\theta,
\end{equation}
with $\rho_\perp$ the perpendicular resistivity, $\rho_\|$ the longitudinal resistivity, $\mathbf{j}$ the current density vector, $\mathbf{m}$ the unit vector pointing in the magnetization direction, and $\theta$ the angle between $\mathbf{j}$ and $\mathbf{m}$. In other words, the highest resistance values can be observed when the magnetization is parallel to the current direction ($\theta=0^\circ$), while the lowest resistance values are obtained when the two directions are perpendicular to each other ($\theta=90^\circ$). A maximum 2\% of the total resistance difference for $\theta=0^\circ$ and $\theta=90^\circ$ was found in our device at 300~K (with an applied field fixed at +1000~Oe). 

Figure \ref{fig:dcexp}(b) shows the AMR curves of the disk for $\theta=0^\circ$ at three representative temperatures: 300~K (top, green), 100~K (middle, orange), and 40~K (bottom, blue). The curves have been vertically shifted for clarity. At 300~K, the change in resistance from 0~Oe to +1000~Oe is in the range of 1\%, corresponding to half of the maximum resistance change (2\%) [Fig. \ref{fig:dcexp}(b)]. This suggests that the same amount of spins points parallel and perpendicular to the current at remanence. Interestingly, the measured curves overlap with each other in the forward and backward half of the loop. Both observations agree with the behavior expected for a vortex state in the disk \cite{Kasai2006}. Moreover, nucleation (annihilation) of the vortex core can be identified as a step-like resistance change in the AMR curve for the forward half of the measurement around +200 ($-600$)~Oe, indicated by I (II) in Fig. \ref{fig:dcexp}(b). Symmetric steps at $-200$ Oe [point I' in Fig. \ref{fig:dcexp}(b)] and +600 Oe [point II' in Fig. \ref{fig:dcexp}(b)] were observed in the backward half of the measurement. All these features indicate a regular vortex-nucleation-annihilation process in the disk \cite{Kasai2006,Sushruth2016, Sushruth2017,Cui2015}, confirming the high quality of our structure. 

We also studied the energy barrier for nucleation and annihilation by measuring the nucleation field $H_\textrm{n}$ and the annihilation field $H_\textrm{an}$ as a function of temperature [Fig. \ref{fig:dcexp}(c)]. In Fig.~\ref{fig:dcexp}(c), we analyzed the average of the absolute value of the up-sweep and down-sweep loops to reduce the uncertainty. 
Different temperature trends of the nucleation and annihilation fields are reported in the literature. For instance, previous work by Mihajlovic et al. demonstrated a peak in nucleation and annihilation fields at low temperature \cite{Mihajlovic2010}, while our results show a monotonic decrease of the annihilation field and a monotonic increase of the nucleation field as the temperature increases in agreement with the behavior found by \v{S}\v{c}epka et al. \cite{Scepka2015}. These differences could be due to the different size, shape, and defects of the samples inevitably occurring during the fabrication process.

At $T = 100$~K, an additional resistance drop is observed around 200 Oe [point III in Fig. \ref{fig:dcexp}(b), orange curve]. The resistance then jumps back up after a few Oe to the reversible behavior characteristic of a magnetic vortex, similar to the results at 300~K. Interestingly, the shape of the AMR curve changes more drastically when the temperature is reduced to 40~K [Fig. \ref{fig:dcexp}(b), blue curve]. After the nucleation (the step-like decrease in resistance indicated by I) around 200~Oe, the resistance rises again as the field decreases. This is a counter-intuitive behavior, since it indicates there is an increase in the magnetization parallel to the field and current. However, we would expect the magnetization to become randomly oriented as the magnetic field approaches to 0, or to have an evenly distributed amount of spins in a flux closure structure such as a magnetic vortex, which would have the effect of reducing the resistance down. The resistance jumps back to the reversible curve corresponding to a magnetic vortex [point IV in Fig. \ref{fig:dcexp}(b), blue curve] when a reversal field of $H_{dep} =-50$ Oe is applied [indicated in Fig. \ref{fig:dcexp}(b)]. The change in the shape of the AMR curve indicates that the magnetic configuration of the disk also changes, eventually reaching a different \textcolor{black}{
equlibrium} state at low temperatures. However, the exact \textcolor{black}{
equlibrium} state has not been determined, and other experiments, such as low temperature MFM, would be needed to characterize the configuration.

To further support our interpretation of the experimental data, we carried out micromagnetic simulations using the LLG micromagnetics simulator \footnote{M.R. Scheinfein, http://llgmicro.home.mindspring.com (1997).}. For this purpose, we simulated a Py disk of 1~$\mu$m diameter and 50 nm thickness, with a cell size of $4\times 4 \times 50$~nm$^3$. Standard parameters for Py (saturation magnetization $M_S=800$ emu/cm$^3$, exchange stiffness constant $A=1.05$ $\mu$erg/cm$^3$) were used in the simulation. The damping parameter was set to 1 to reach the equilibrium state faster. We also simulate the current distribution using the LLG micromagnetics simulator. Finally, we obtain the resistance as the sum of the product of the current and the magnetization in each cell [$R\propto\sum (\mathbf{j}_i\cdot\mathbf{m}_i)$, from Eq. \eqref{amr}]. To account for a temperature dependence of the magnetization in the simulations, we used a Langevin random field that simulates a temperature of 100 K. The discrepancy in the exact temperature values can be produced by different experimental and simulation parameters, by the defects inside the material and by the fact that the samples are not perfectly shaped. 
\begin{figure}
\includegraphics[width=\columnwidth]{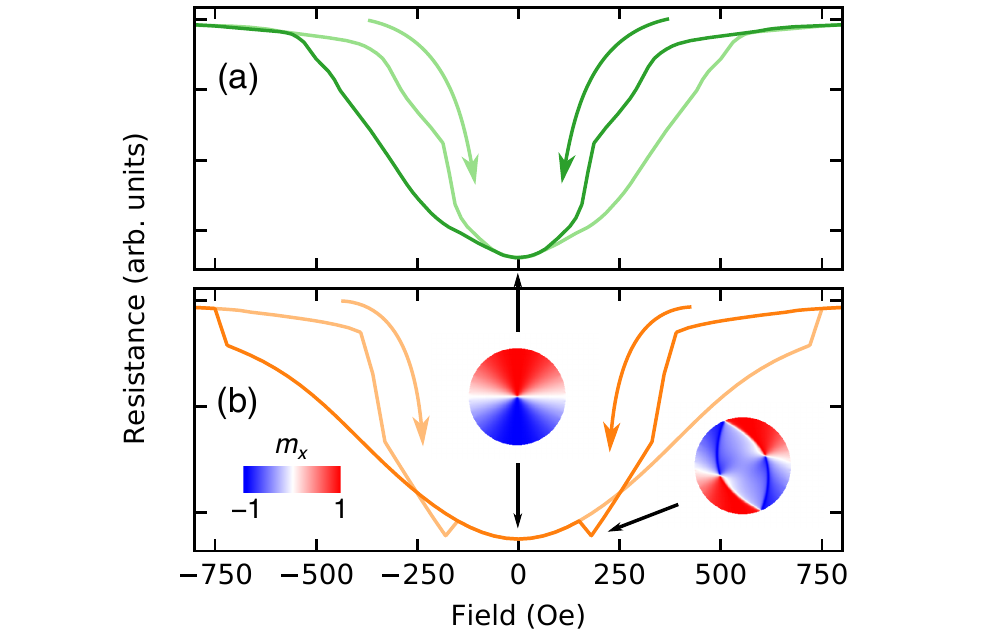}
\caption{Micromagnetic simulations showing the AMR of a 1-$\mu m$-diameter disk with (a) and without (b) the Langevin term, which simulates the effects of temperature in the system. The colored arrows show the direction of the field sweep, while insets show the magnetization state at the regions indicated by the arrows. \label{fig:sim}}
\end{figure}

In Fig.~\ref{fig:sim} we compare the results of the simulations with and without taking into account effects of temperature; i.e., with and without the Langevin term. In the simulations without the Langevin term, we obtained an AMR curve with a peak at +200 Oe [see Fig. \ref{fig:sim}(b)], similar to the peak found in the experimental data measured at 100 K [point III in Fig. \ref{fig:dcexp}(b)]. The magnetic configuration in this case consists of two vortices on opposite sites of the disk, coming from an S-type nucleation. On the other hand, when we run the simulation with a Langevin term accounting for temperature in the system, the AMR changes shape, as shown in Fig. \ref{fig:sim}(a). In this case, there is no minor peak in the AMR, and the curve resembles the experimental curve at 300 K [see Fig. \ref{fig:dcexp}(b)]. Before the field is reduced to 0, the magnetic vortex is nucleated [the magnetic configuration of the vortex at 0 field is shown in Fig. \ref{fig:sim}]. As the field is further swept, the core is shifted away from the center of the disk until it is annihilated. We find that the annihilation field is also temperature dependent in the simulations: when we take the Langevin term into account (elevated temperature), the annihilation field is close to +500 Oe, while it is close to +750 Oe when we do not consider the Langevin term in the simulations (zero temperature). These results agree well with the experimental observations. Our experimental and simulation data indicate that, at 100 K, the magnetic state can change between a single vortex and two vortices. The transition between these two states is not trivial, and our results show that AMR measurements are a potential way to detect them. However, we were not able to reproduce the experimental AMR curve observed at temperatures below 40 K with micromagnetic simulations. Nonetheless, our results indicate that intermediate states appear as the temperature is reduced, probably due to pinning effects around defects. In this scenario, temperature fluctuations may help to overcome pinning barriers, hence changing the energy landscape and the nucleation process.

\section{Magnetization Dynamics}\label{sec:rf}

In the first part of the manuscript, we have laid out the groundwork for the rf measurements, which will be discussed next. We confirmed that occurrence of magnetic vortices is accompanied with characteristic features in the magnetic transport measurements, which was further correlated with micromagnetic simulations.

\subsection{Experimental setup for rf measurements}\label{sec:rfsetup}

\begin{figure}
\includegraphics[width=\columnwidth]{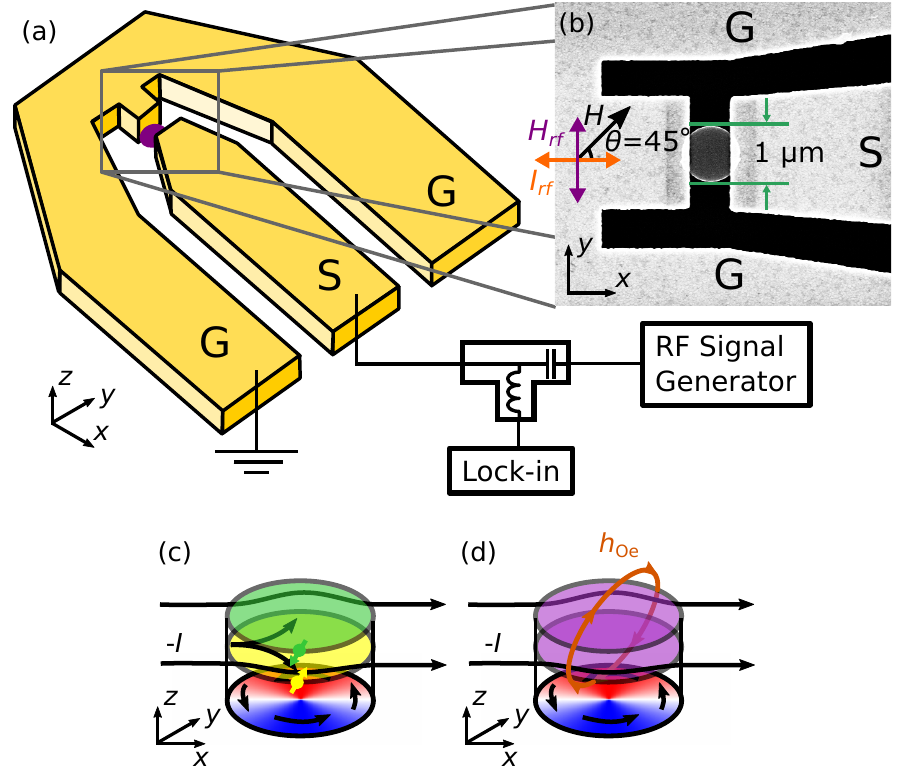}
\caption{\textcolor{black}{(a)} Schematics of the CPW (yellow) with the signal (S) and ground (G) lines shorted, with a Py/Pt disk (blue) in the S line.  A bias-T was used to split the dc and rf components of the signal, which were connected to a lock-in amplifier and an rf signal generator, respectively. \textcolor{black}{(b)} SEM image of the 1-$\mu$m-diameter disk (dark grey), in the S line of the patterned CPW (white); the dark area is the Si substrate. The field is applied at an angle $\theta = 45^\circ$ with respect to the current. \textcolor{black}{(c) Spin-Hall effect in the Pt layer generates spin accumulation at the surfaces (yellow and green), that can exert a torque on the magnetization in the bottom layer and in turn lead to the onset of dynamics. (d) As current passes through Pt, an Oersted field is generated (brown) driving magnetization dynamics in the bottom layer.} \label{fig:rfsetup}}
\end{figure}

With the goal of exciting low-frequency dynamics by spin torque, we fabricated a single microdisk in a 700-nm gap of the signal line of a coplanar waveguide (CPW) of 2-$\mu$m width, 5-nm Ti/100-nm Au, see Fig. \ref{fig:rfsetup}(a). A lift-off process was used to pattern the disks, in combination with the deposition of a Py(35~nm)/Pt(5~nm) bilayer using electron-beam evaporation in one single step. A SEM image of the sample is shown in Fig.~\ref{fig:rfsetup}(b). 
The heavy metal Pt serves as a spin current source: The alternating microwave current is converted into a spin-polarized current via the spin Hall Effect (SHE) [Fig. \ref{fig:rfsetup}(c)], which in turn drives the magnetization by spin torque. In addition, there is a contribution from the Oersted field created by the microwave current [Fig. \ref{fig:rfsetup}(d)]. The vortex dynamics are detected by a rectified voltage produced by a time-averaged mixing of the anisotropic magnetoresistance with the microwave current. The same idea has been used to study high-frequency dynamics in ferromagnet/heavy metal bilayer microstructures and is known as spin torque ferromagnetic resonance (ST-FMR)  \cite{Liu2011}. However, in our inhomogeneous magnetized sample containing a magnetic vortex, a signal only arises when the magnetic core is off-centered  and the symmetric contributions are broken. In order to enhance the spin-orbit torque, which arises in the vicinity of the Py/Pt interface, we reduced the Py thickness from 50 nm to 35 nm. This thickness, however, is still large enough to ensure the formation of a magnetic vortex. While the different thickness changes the exact values of nucleation and annihilation fields, the dynamics once the vortex is generated are expected to be comparable.  
We note that the additional Pt layer might induce surface anisotropy, which in turn, can alter the nucleation/annihilation fields of the magnetic vortex \cite{Im2012,Luo2014,Luo2015}. However, we emphasize that the AMR data of this bilayer sample [Fig.~\ref{fig:rfresult}(a)] shows a very similar behavior as the single Py sample (without Pt) shown in Fig.~\ref{fig:dcexp}(b) for 100~K and the simulated AMR [Fig.~\ref{fig:sim}(b)]. This indicates that even if the exact nucleation and annihilation process and fields might differ due to different thicknesses and possible presence of surface anisotropy, the general trend is similar.

In order to excite the gyrotropic motion and to measure the rectified voltage, a microwave signal is passed through the CPW via a bias tee. The sample connections are made with a probe that matches the size of the lithographically patterned CPW and the rectified voltage is measured by a lock-in amplifier (see Fig. \ref{fig:rfsetup}). For the dynamics measurements, the microwave signal is modulated at 1.4 kHz. The measurement proceeds as follows: a particular frequency value between 200 MHz and 700 MHz is set, and then the field is swept from $-1000$ Oe to 1000 Oe in 10 Oe steps; at each field step, the voltage is read from the lock-in amplifier. The field was applied at $\theta=45^\circ$ with respect to the direction of the CPW line. This was done to maximize the signal when the sample is saturated in analogy to ST-FMR measurements. In this configuration, the signal is maximized since the change in magnetoresistance, ultimately producing the measured voltage response, is maximum. 
After each field step, the frequency was changed in 10 MHz steps to record the full spectrum from 200 MHz to 700 MHz.

\subsection{Results of rf measurements}\label{sec:rfresults}

\begin{figure}
\includegraphics[width=\columnwidth]{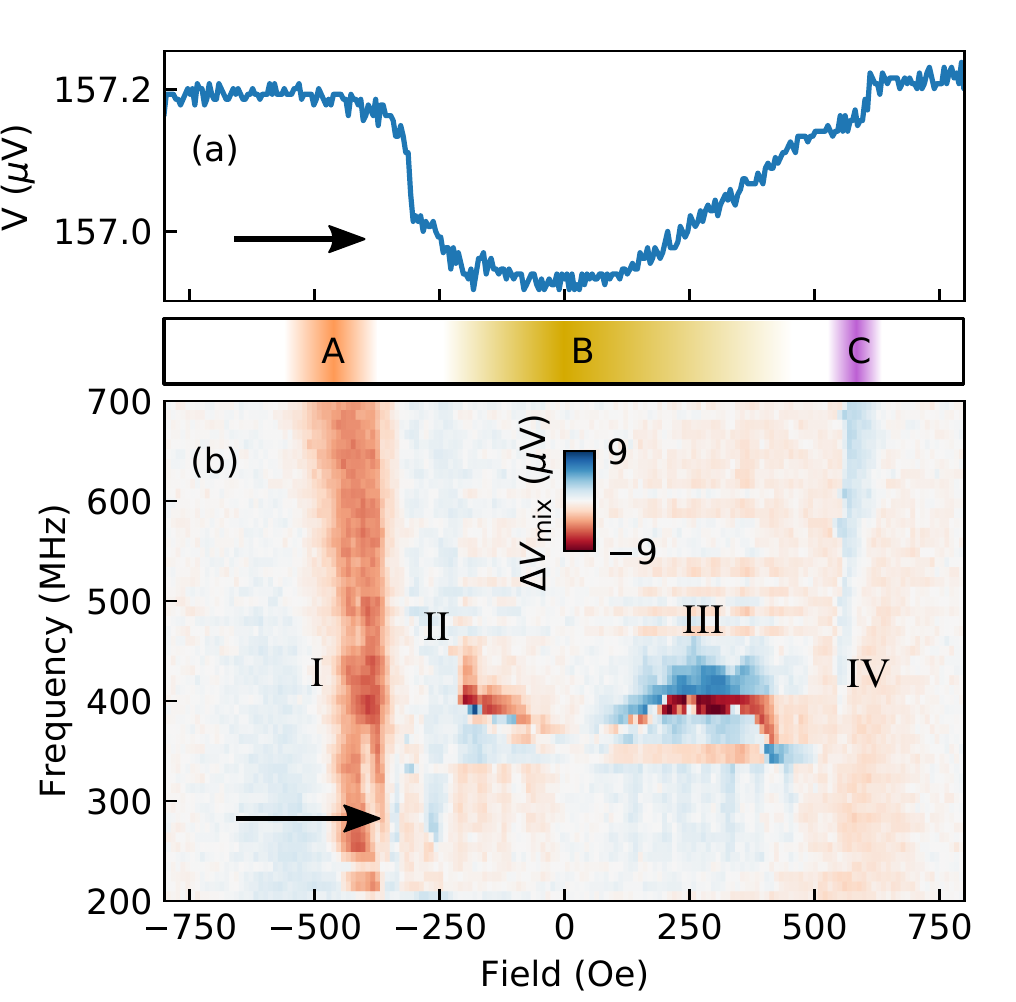}
\caption{(a) Two-terminal AMR measured at 300 K sweeping the field in the same direction. Nucleation occurs around 300 Oe, and annihilation around 600 Oe. (b) Excitation spectrum showing the normalized mixed voltage in the color scale as a function of field and frequency. Black arrows show the direction of the field sweep. As the field is increased, a broadband mode is first observed (I), then the gyrotropic mode starts at around 400 MHz (II). The gyrotropic mode is symmetric around 0 Oe until $\approx 300$ MHz (III). Once the gyrotropic mode disappears, a smaller broadband mode appears (IV). The central pannel shows the dominant excitations depending on the field: (A, orange) first broadband response, (B, yellow) gyrotropic motion, and (C, purple) second broadband response. \label{fig:rfresult}}
\end{figure}

Figure \ref{fig:rfresult}(b) shows a representative spectrum of a 1 $\mu$m-diameter disk excited at 0 dBm and measured at 300 K, plotted as a color map. In this map, red (blue) represents a low (high) voltage signal. 
As we sweep the field up from a negative saturation value of $-1.5$ kOe (not shown), a broadband response appears at a negative field of $-400$ Oe [I in Fig. \ref{fig:rfsetup}(b)]. This band has a width of approximately 50 Oe (as indicated by the A, orange region in the central panel) and it can be observed from 200 MHz (which is the lower limit of our experimental setup) up to several GHz (not shown).
At low fields (between $\approx-250$ Oe and $+500$ Oe), we can identify a signal around 400 MHz that corresponds to the gyrotropic motion of the magnetic vortex core around its equilibrium position, in agreement with its theoretical value \cite{Guslienko2006} taking into account a value for the saturation magnetization $M_\mathrm{S}=750$ emu/cm$^3$.  
The gyrotropic mode starts at $\approx-250$~Oe with a frequency of 400~MHz as a magnetic vortex is generated. Since the vortex core is not at the center of the disk, there is a net magnetization, and hence there is a measurable signal. As the field is reduced, the frequency decreases to the minimum value of $\approx 375$ MHz at 0 Oe, at which point the mixed voltage is zero since there are cancelling contributions from spins pointing in opposite directions; i.e., there is no net magnetization when the vortex is at the center. Then, the frequency increases again up to 400 MHz as the field is increased to $\approx+250$ Oe and the vortex moves out of the disk center. By continuing increasing the magnetic field, the gyrotropic frequency goes down again until a field value of +400 Oe is reached, at which point the mode disappears (B, yellow region in the central panel). At a field value of $\approx+600$ Oe, a broadband mode starting at 400 MHz appears (C, purple region in central panel). The asymmetric behavior of the gyrotropic mode with respect to field is due to the different magnetization processes involved in nucleation (negative fields) and annihilation (positive fields), as observed in the asymmetry of the AMR data [Fig. \ref{fig:dcexp}(b) and Fig. \ref{fig:rfresult}(a)].

\textcolor{black}{
The broadband response (marked as A and C in the center panel of Fig.~\ref{fig:rfresult}) extends beyond the shown range of 200 MHz to 700 MHz}, and can even be detected at higher frequencies above 5 GHz (not shown). By comparing the field values at which these modes appear with the measured AMR [Fig.~\ref{fig:rfresult}(a)], it becomes evident that the spin-torque dynamics indicate the onset of vortex dynamics even before the vortex is nucleated. In this transition regime between single domain and vortex state, the microwave/spin-torque drive can effectively couple to the magnetic moments and start the precession, which is indicated by the broadband response. As the magnetic field is swept and changes polarity, the vortex first nucleates and then eventually annihilates. Afterwards, a second broad response appears (regime IV). This second broadband response persists for a few tens of Oe as the magnitude of the field is increased. Again, by a comparison with the \textcolor{black}{dc} AMR data, the second band can be correlated with the annihilation of the vortex. In this context, we note that telegraph-like noise\textcolor{black}{,} appearing when a system keeps hopping between two states\textcolor{black}{,} has been previously shown to produce low-frequency signals \cite{Burgess2013,Jenkins2019}.
 
 To summarize these observations, the nucleation (annihilation) fields can be determined by the field at which the broadband response disappears (reappears) in the \textcolor{black}{
 rf} measurements. Furthermore, our results suggest that the \textcolor{black}{
 dynamic} approach is more sensitive to the vortex formation process than the corresponding static AMR measurements. 
The exact quantification of the spin-torque and Oersted field contribution requires a refined theoretical model that is \textcolor{black}{
outside} the scope of the current work.

\begin{figure}
\centering
\includegraphics{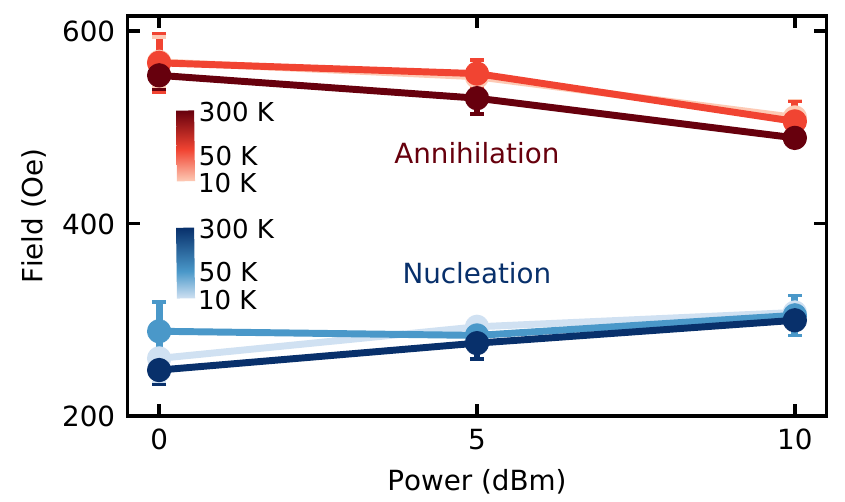}
\caption{Dependence of nucleation (blue curves at the bottom) and annihilation (red curves on top) fields \textcolor{black}{
on the applied rf power}, measured at temperatures of 10~K, 50~K and 300~K (bright to dark colors). The curves show an increase in the nucleation and a decrease in the annihilation as the power is increased. The values and associated error bars were obtained from the average of the different extracted values at each frequency, as shown in Fig. \ref{fig:rfresult}. \label{fig:nucan}}
\end{figure}

To further study the spin-torque driven gyrotropic motion, power dependent measurements (+5 dBm and +10 dBm) at low temperatures (10 K and 50 K) were carried out.

Figure \ref{fig:nucan} compares the variation of the nucleation and annihilation fields for the three different measured microwave powers and temperatures. The nucleation field values are obtained from the spectra by determining the mean field at which the first broadband response [I in Fig. \ref{fig:rfresult}(b)] disappears for different frequencies. The annihilation field is obtained as the mean at which the second broadband [IV in Fig. \ref{fig:rfresult}] starts for different frequencies. As it is apparent from Fig.~\ref{fig:nucan}, the nucleation field increases with power, whereas the annihilation field decreases with increasing power.
Comparing with the experimentally acquired trends obtained from temperature-dependent AMR measurements, shown in Fig. \ref{fig:dcexp}(c), this suggests that heat produced by the $rf$ excitation is at least partially responsible for the reduction (increase) of the annihilation (nucleation) field. Heat development may also explain the discrepancies in the exact values of the nucleation and annihilation field extracted from the AMR curve [Fig.~\ref{fig:rfresult}(a)] and from the onset of the broadband response in the spin-torque driven dynamics. Another possible mechanism for the change of the nucleation and annihilation energy barriers is the forced oscillation of the magnetic moments. In that regard, this would be similar to the broadband response of microwave assisted switching with granular magnetic media \cite{Lu2013}.

\textcolor{black}{
The smaller change in the nucleation and annihilation fields obtained from the dynamic data (Fig.~\ref{fig:nucan}) compared to the dc AMR data [Fig.~\ref{fig:dcexp}(c)] points to a more similar vortex formation/annihilation process across temperatures. Since the magnetic configuration is unknown, f}urther measurements to better understand this behavior would be needed. For instance, imaging techniques such as magnetic force microscopy, scanning transmission X-ray microscopy and Brillouin light scattering would be powerful tools to unveil the details of both the quasi-static magnetic state and of the dynamics in our system. Such techniques have already been successfully used to determine the magnetic state of patterned structures \cite{Shinjo2000}, unveiling non-adiabatic torques in a magnetic vortex \cite{Bisig2016} and in the context of spin-orbit-torque excitation of planar structures \cite{Baumgartner2017,Haidar2019}.

\section{Conclusions}\label{sec:conclusion}

In summary, we have demonstrated that the concept of spin-torque ferromagnetic resonance can be extended to the excitation and detection of the gyrotropic motion in a single magnetic disk. However, the determination of the exact contribution of the spin-torque and the Oersted field requires a non-macrospin theoretical expression and an improved signal-to-noise ratio. We presented a comprehensive study of the dc and low-frequency rf response at different base temperatures and power levels. Using the dc measurements of the anisotropic magnetoresistance in combination with micromagnetic simulations we characterized the dependence of the nucleation and annihilation fields with temperature, and observed that different magnetic configurations occur at lower temperatures. Furthermore, we found that a broadband response in the dynamic measurements occurs before and after the vortex is created/destroyed. Our results are the first steps towards the integration of low-frequency dynamics in spin-torque devices.

\section*{Acknowledgments}

Work at Argonne was supported by the U.S. Department of Energy (DOE), Office of Science, Materials Science and Engineering Division. Work at Delaware was supported by the National Science Foundation under Grant No. 1833000. The use of the Center for Nanoscale Materials is supported by DOE-BES, under Contract No. DE-AC02-06CH11357.

\section*{Data Availability}
The data that support the findings of this study are available from the corresponding author upon reasonable request.


\begin{thebibliography}{43}%
\makeatletter
\providecommand \@ifxundefined [1]{%
 \@ifx{#1\undefined}
}%
\providecommand \@ifnum [1]{%
 \ifnum #1\expandafter \@firstoftwo
 \else \expandafter \@secondoftwo
 \fi
}%
\providecommand \@ifx [1]{%
 \ifx #1\expandafter \@firstoftwo
 \else \expandafter \@secondoftwo
 \fi
}%
\providecommand \natexlab [1]{#1}%
\providecommand \enquote  [1]{``#1''}%
\providecommand \bibnamefont  [1]{#1}%
\providecommand \bibfnamefont [1]{#1}%
\providecommand \citenamefont [1]{#1}%
\providecommand \href@noop [0]{\@secondoftwo}%
\providecommand \href [0]{\begingroup \@sanitize@url \@href}%
\providecommand \@href[1]{\@@startlink{#1}\@@href}%
\providecommand \@@href[1]{\endgroup#1\@@endlink}%
\providecommand \@sanitize@url [0]{\catcode `\\12\catcode `\$12\catcode
  `\&12\catcode `\#12\catcode `\^12\catcode `\_12\catcode `\%12\relax}%
\providecommand \@@startlink[1]{}%
\providecommand \@@endlink[0]{}%
\providecommand \url  [0]{\begingroup\@sanitize@url \@url }%
\providecommand \@url [1]{\endgroup\@href {#1}{\urlprefix }}%
\providecommand \urlprefix  [0]{URL }%
\providecommand \Eprint [0]{\href }%
\providecommand \doibase [0]{http://dx.doi.org/}%
\providecommand \selectlanguage [0]{\@gobble}%
\providecommand \bibinfo  [0]{\@secondoftwo}%
\providecommand \bibfield  [0]{\@secondoftwo}%
\providecommand \translation [1]{[#1]}%
\providecommand \BibitemOpen [0]{}%
\providecommand \bibitemStop [0]{}%
\providecommand \bibitemNoStop [0]{.\EOS\space}%
\providecommand \EOS [0]{\spacefactor3000\relax}%
\providecommand \BibitemShut  [1]{\csname bibitem#1\endcsname}%
\let\auto@bib@innerbib\@empty
\bibitem [{\citenamefont {Liu}\ \emph {et~al.}(2012{\natexlab{a}})\citenamefont
  {Liu}, \citenamefont {Pai}, \citenamefont {Li}, \citenamefont {Tseng},
  \citenamefont {Ralph},\ and\ \citenamefont {Buhrman}}]{Liu2012}%
  \BibitemOpen
  \bibfield  {author} {\bibinfo {author} {\bibfnamefont {L.}~\bibnamefont
  {Liu}}, \bibinfo {author} {\bibfnamefont {C.-F.}\ \bibnamefont {Pai}},
  \bibinfo {author} {\bibfnamefont {Y.}~\bibnamefont {Li}}, \bibinfo {author}
  {\bibfnamefont {H.~W.}\ \bibnamefont {Tseng}}, \bibinfo {author}
  {\bibfnamefont {D.~C.}\ \bibnamefont {Ralph}}, \ and\ \bibinfo {author}
  {\bibfnamefont {R.~A.}\ \bibnamefont {Buhrman}},\ }\href {\doibase
  10.1126/science.1218197} {\bibfield  {journal} {\bibinfo  {journal}
  {Science}\ }\textbf {\bibinfo {volume} {336}},\ \bibinfo {pages} {555}
  (\bibinfo {year} {2012}{\natexlab{a}})}\BibitemShut {NoStop}%
\bibitem [{\citenamefont {Liu}\ \emph {et~al.}(2012{\natexlab{b}})\citenamefont
  {Liu}, \citenamefont {Lee}, \citenamefont {Gudmundsen}, \citenamefont
  {Ralph},\ and\ \citenamefont {Buhrman}}]{Liu2012a}%
  \BibitemOpen
  \bibfield  {author} {\bibinfo {author} {\bibfnamefont {L.}~\bibnamefont
  {Liu}}, \bibinfo {author} {\bibfnamefont {O.~J.}\ \bibnamefont {Lee}},
  \bibinfo {author} {\bibfnamefont {T.~J.}\ \bibnamefont {Gudmundsen}},
  \bibinfo {author} {\bibfnamefont {D.~C.}\ \bibnamefont {Ralph}}, \ and\
  \bibinfo {author} {\bibfnamefont {R.~A.}\ \bibnamefont {Buhrman}},\ }\href
  {\doibase 10.1103/PhysRevLett.109.096602} {\bibfield  {journal} {\bibinfo
  {journal} {Phys. Rev. Lett.}\ }\textbf {\bibinfo {volume} {109}},\ \bibinfo
  {pages} {096602} (\bibinfo {year} {2012}{\natexlab{b}})}\BibitemShut
  {NoStop}%
\bibitem [{\citenamefont {Fukami}\ \emph {et~al.}(2016)\citenamefont {Fukami},
  \citenamefont {Zhang}, \citenamefont {DuttaGupta}, \citenamefont {Kurenkov},\
  and\ \citenamefont {Ohno}}]{Fukami2016}%
  \BibitemOpen
  \bibfield  {author} {\bibinfo {author} {\bibfnamefont {S.}~\bibnamefont
  {Fukami}}, \bibinfo {author} {\bibfnamefont {C.}~\bibnamefont {Zhang}},
  \bibinfo {author} {\bibfnamefont {S.}~\bibnamefont {DuttaGupta}}, \bibinfo
  {author} {\bibfnamefont {A.}~\bibnamefont {Kurenkov}}, \ and\ \bibinfo
  {author} {\bibfnamefont {H.}~\bibnamefont {Ohno}},\ }\href {\doibase
  10.1038/nmat4566} {\bibfield  {journal} {\bibinfo  {journal} {Nat. Mater.}\
  }\textbf {\bibinfo {volume} {15}},\ \bibinfo {pages} {535} (\bibinfo {year}
  {2016})}\BibitemShut {NoStop}%
\bibitem [{\citenamefont {Liu}\ \emph {et~al.}(2011)\citenamefont {Liu},
  \citenamefont {Moriyama}, \citenamefont {Ralph},\ and\ \citenamefont
  {Buhrman}}]{Liu2011}%
  \BibitemOpen
  \bibfield  {author} {\bibinfo {author} {\bibfnamefont {L.}~\bibnamefont
  {Liu}}, \bibinfo {author} {\bibfnamefont {T.}~\bibnamefont {Moriyama}},
  \bibinfo {author} {\bibfnamefont {D.~C.}\ \bibnamefont {Ralph}}, \ and\
  \bibinfo {author} {\bibfnamefont {R.~A.}\ \bibnamefont {Buhrman}},\ }\href
  {\doibase 10.1103/PhysRevLett.106.036601} {\bibfield  {journal} {\bibinfo
  {journal} {Phys. Rev. Lett.}\ }\textbf {\bibinfo {volume} {106}},\ \bibinfo
  {pages} {036601} (\bibinfo {year} {2011})}\BibitemShut {NoStop}%
\bibitem [{\citenamefont {Parkin}, \citenamefont {Hayashi},\ and\ \citenamefont
  {Thomas}(2008)}]{Parkin2008}%
  \BibitemOpen
  \bibfield  {author} {\bibinfo {author} {\bibfnamefont {S.~S.~P.}\
  \bibnamefont {Parkin}}, \bibinfo {author} {\bibfnamefont {M.}~\bibnamefont
  {Hayashi}}, \ and\ \bibinfo {author} {\bibfnamefont {L.}~\bibnamefont
  {Thomas}},\ }\href {\doibase 10.1126/science.1145799} {\bibfield  {journal}
  {\bibinfo  {journal} {Science}\ }\textbf {\bibinfo {volume} {320}},\ \bibinfo
  {pages} {190} (\bibinfo {year} {2008})}\BibitemShut {NoStop}%
\bibitem [{\citenamefont {Parkin}\ and\ \citenamefont
  {Yang}(2015)}]{Parkin2015}%
  \BibitemOpen
  \bibfield  {author} {\bibinfo {author} {\bibfnamefont {S.}~\bibnamefont
  {Parkin}}\ and\ \bibinfo {author} {\bibfnamefont {S.-H.}\ \bibnamefont
  {Yang}},\ }\href {\doibase 10.1038/nnano.2015.41} {\bibfield  {journal}
  {\bibinfo  {journal} {Nat. Nanotechnol.}\ }\textbf {\bibinfo {volume} {10}},\
  \bibinfo {pages} {195} (\bibinfo {year} {2015})}\BibitemShut {NoStop}%
\bibitem [{\citenamefont {Liu}, \citenamefont {Lim},\ and\ \citenamefont
  {Urazhdin}(2013)}]{Liu2013}%
  \BibitemOpen
  \bibfield  {author} {\bibinfo {author} {\bibfnamefont {R.~H.}\ \bibnamefont
  {Liu}}, \bibinfo {author} {\bibfnamefont {W.~L.}\ \bibnamefont {Lim}}, \ and\
  \bibinfo {author} {\bibfnamefont {S.}~\bibnamefont {Urazhdin}},\ }\href
  {\doibase 10.1103/PhysRevLett.110.147601} {\bibfield  {journal} {\bibinfo
  {journal} {Phys. Rev. Lett.}\ }\textbf {\bibinfo {volume} {110}},\ \bibinfo
  {pages} {147601} (\bibinfo {year} {2013})}\BibitemShut {NoStop}%
\bibitem [{\citenamefont {Demidov}\ \emph
  {et~al.}(2014{\natexlab{a}})\citenamefont {Demidov}, \citenamefont {Ulrichs},
  \citenamefont {Gurevich}, \citenamefont {Demokritov}, \citenamefont
  {Tiberkevich}, \citenamefont {Slavin}, \citenamefont {Zholud},\ and\
  \citenamefont {Urazhdin}}]{Demidov2014}%
  \BibitemOpen
  \bibfield  {author} {\bibinfo {author} {\bibfnamefont {V.~E.}\ \bibnamefont
  {Demidov}}, \bibinfo {author} {\bibfnamefont {H.}~\bibnamefont {Ulrichs}},
  \bibinfo {author} {\bibfnamefont {S.~V.}\ \bibnamefont {Gurevich}}, \bibinfo
  {author} {\bibfnamefont {S.~O.}\ \bibnamefont {Demokritov}}, \bibinfo
  {author} {\bibfnamefont {V.~S.}\ \bibnamefont {Tiberkevich}}, \bibinfo
  {author} {\bibfnamefont {A.~N.}\ \bibnamefont {Slavin}}, \bibinfo {author}
  {\bibfnamefont {A.}~\bibnamefont {Zholud}}, \ and\ \bibinfo {author}
  {\bibfnamefont {S.}~\bibnamefont {Urazhdin}},\ }\href {\doibase
  10.1038/ncomms4179} {\bibfield  {journal} {\bibinfo  {journal} {Nat.
  Commun.}\ }\textbf {\bibinfo {volume} {5}},\ \bibinfo {pages} {3179}
  (\bibinfo {year} {2014}{\natexlab{a}})}\BibitemShut {NoStop}%
\bibitem [{\citenamefont {Demidov}\ \emph
  {et~al.}(2014{\natexlab{b}})\citenamefont {Demidov}, \citenamefont
  {Urazhdin}, \citenamefont {Zholud}, \citenamefont {Sadovnikov},\ and\
  \citenamefont {Demokritov}}]{Demidov2014a}%
  \BibitemOpen
  \bibfield  {author} {\bibinfo {author} {\bibfnamefont {V.~E.}\ \bibnamefont
  {Demidov}}, \bibinfo {author} {\bibfnamefont {S.}~\bibnamefont {Urazhdin}},
  \bibinfo {author} {\bibfnamefont {A.}~\bibnamefont {Zholud}}, \bibinfo
  {author} {\bibfnamefont {A.~V.}\ \bibnamefont {Sadovnikov}}, \ and\ \bibinfo
  {author} {\bibfnamefont {S.~O.}\ \bibnamefont {Demokritov}},\ }\href
  {\doibase 10.1063/1.4901027} {\bibfield  {journal} {\bibinfo  {journal}
  {Appl. Phys. Lett.}\ }\textbf {\bibinfo {volume} {105}},\ \bibinfo {pages}
  {172410} (\bibinfo {year} {2014}{\natexlab{b}})}\BibitemShut {NoStop}%
\bibitem [{\citenamefont {Locatelli}, \citenamefont {Cros},\ and\ \citenamefont
  {Grollier}(2014)}]{Locatelli2014}%
  \BibitemOpen
  \bibfield  {author} {\bibinfo {author} {\bibfnamefont {N.}~\bibnamefont
  {Locatelli}}, \bibinfo {author} {\bibfnamefont {V.}~\bibnamefont {Cros}}, \
  and\ \bibinfo {author} {\bibfnamefont {J.}~\bibnamefont {Grollier}},\ }\href
  {\doibase 10.1038/nmat3823} {\bibfield  {journal} {\bibinfo  {journal} {Nat.
  Mater.}\ }\textbf {\bibinfo {volume} {13}},\ \bibinfo {pages} {11} (\bibinfo
  {year} {2014})}\BibitemShut {NoStop}%
\bibitem [{\citenamefont {Brataas}, \citenamefont {Kent},\ and\ \citenamefont
  {Ohno}(2012)}]{Brataas2012}%
  \BibitemOpen
  \bibfield  {author} {\bibinfo {author} {\bibfnamefont {A.}~\bibnamefont
  {Brataas}}, \bibinfo {author} {\bibfnamefont {A.~D.}\ \bibnamefont {Kent}}, \
  and\ \bibinfo {author} {\bibfnamefont {H.}~\bibnamefont {Ohno}},\ }\href
  {\doibase 10.1038/nmat3311} {\bibfield  {journal} {\bibinfo  {journal} {Nat.
  Mater.}\ }\textbf {\bibinfo {volume} {11}},\ \bibinfo {pages} {372} (\bibinfo
  {year} {2012})}\BibitemShut {NoStop}%
\bibitem [{\citenamefont {Baumgartner}\ \emph {et~al.}(2017)\citenamefont
  {Baumgartner}, \citenamefont {Garello}, \citenamefont {Mendil}, \citenamefont
  {Avci}, \citenamefont {Grimaldi}, \citenamefont {Murer}, \citenamefont
  {Feng}, \citenamefont {Gabureac}, \citenamefont {Stamm}, \citenamefont
  {Acremann}, \citenamefont {Finizio}, \citenamefont {Wintz}, \citenamefont
  {Raabe},\ and\ \citenamefont {Gambardella}}]{Baumgartner2017}%
  \BibitemOpen
  \bibfield  {author} {\bibinfo {author} {\bibfnamefont {M.}~\bibnamefont
  {Baumgartner}}, \bibinfo {author} {\bibfnamefont {K.}~\bibnamefont
  {Garello}}, \bibinfo {author} {\bibfnamefont {J.}~\bibnamefont {Mendil}},
  \bibinfo {author} {\bibfnamefont {C.~O.}\ \bibnamefont {Avci}}, \bibinfo
  {author} {\bibfnamefont {E.}~\bibnamefont {Grimaldi}}, \bibinfo {author}
  {\bibfnamefont {C.}~\bibnamefont {Murer}}, \bibinfo {author} {\bibfnamefont
  {J.}~\bibnamefont {Feng}}, \bibinfo {author} {\bibfnamefont {M.}~\bibnamefont
  {Gabureac}}, \bibinfo {author} {\bibfnamefont {C.}~\bibnamefont {Stamm}},
  \bibinfo {author} {\bibfnamefont {Y.}~\bibnamefont {Acremann}}, \bibinfo
  {author} {\bibfnamefont {S.}~\bibnamefont {Finizio}}, \bibinfo {author}
  {\bibfnamefont {S.}~\bibnamefont {Wintz}}, \bibinfo {author} {\bibfnamefont
  {J.}~\bibnamefont {Raabe}}, \ and\ \bibinfo {author} {\bibfnamefont
  {P.}~\bibnamefont {Gambardella}},\ }\href {\doibase 10.1038/nnano.2017.151}
  {\bibfield  {journal} {\bibinfo  {journal} {Nat. Nanotechnol.}\ }\textbf
  {\bibinfo {volume} {12}},\ \bibinfo {pages} {980} (\bibinfo {year}
  {2017})}\BibitemShut {NoStop}%
\bibitem [{\citenamefont {Manchon}\ \emph {et~al.}(2019)\citenamefont
  {Manchon}, \citenamefont {{\v{Z}}elezn{\'{y}}}, \citenamefont {Miron},
  \citenamefont {Jungwirth}, \citenamefont {Sinova}, \citenamefont {Thiaville},
  \citenamefont {Garello},\ and\ \citenamefont {Gambardella}}]{Manchon2019}%
  \BibitemOpen
  \bibfield  {author} {\bibinfo {author} {\bibfnamefont {A.}~\bibnamefont
  {Manchon}}, \bibinfo {author} {\bibfnamefont {J.}~\bibnamefont
  {{\v{Z}}elezn{\'{y}}}}, \bibinfo {author} {\bibfnamefont {I.}~\bibnamefont
  {Miron}}, \bibinfo {author} {\bibfnamefont {T.}~\bibnamefont {Jungwirth}},
  \bibinfo {author} {\bibfnamefont {J.}~\bibnamefont {Sinova}}, \bibinfo
  {author} {\bibfnamefont {A.}~\bibnamefont {Thiaville}}, \bibinfo {author}
  {\bibfnamefont {K.}~\bibnamefont {Garello}}, \ and\ \bibinfo {author}
  {\bibfnamefont {P.}~\bibnamefont {Gambardella}},\ }\href {\doibase
  10.1103/RevModPhys.91.035004} {\bibfield  {journal} {\bibinfo  {journal}
  {Rev. Mod. Phys.}\ }\textbf {\bibinfo {volume} {91}},\ \bibinfo {pages}
  {035004} (\bibinfo {year} {2019})}\BibitemShut {NoStop}%
\bibitem [{\citenamefont {Awad}\ \emph {et~al.}(2016)\citenamefont {Awad},
  \citenamefont {D{\"{u}}rrenfeld}, \citenamefont {Houshang}, \citenamefont
  {Dvornik}, \citenamefont {Iacocca}, \citenamefont {Dumas},\ and\
  \citenamefont {{\AA}kerman}}]{Awad2016}%
  \BibitemOpen
  \bibfield  {author} {\bibinfo {author} {\bibfnamefont {A.~A.}\ \bibnamefont
  {Awad}}, \bibinfo {author} {\bibfnamefont {P.}~\bibnamefont
  {D{\"{u}}rrenfeld}}, \bibinfo {author} {\bibfnamefont {A.}~\bibnamefont
  {Houshang}}, \bibinfo {author} {\bibfnamefont {M.}~\bibnamefont {Dvornik}},
  \bibinfo {author} {\bibfnamefont {E.}~\bibnamefont {Iacocca}}, \bibinfo
  {author} {\bibfnamefont {R.~K.}\ \bibnamefont {Dumas}}, \ and\ \bibinfo
  {author} {\bibfnamefont {J.}~\bibnamefont {{\AA}kerman}},\ }\href {\doibase
  10.1038/nphys3927} {\bibfield  {journal} {\bibinfo  {journal} {Nat. Phys.}\
  }\textbf {\bibinfo {volume} {13}},\ \bibinfo {pages} {292} (\bibinfo {year}
  {2016})}\BibitemShut {NoStop}%
\bibitem [{\citenamefont {Woo}\ \emph {et~al.}(2017)\citenamefont {Woo},
  \citenamefont {Song}, \citenamefont {Han}, \citenamefont {Jung},
  \citenamefont {Im}, \citenamefont {Lee}, \citenamefont {Song}, \citenamefont
  {Fischer}, \citenamefont {Hong}, \citenamefont {Choi}, \citenamefont {Min},
  \citenamefont {Koo},\ and\ \citenamefont {Chang}}]{Woo2017}%
  \BibitemOpen
  \bibfield  {author} {\bibinfo {author} {\bibfnamefont {S.}~\bibnamefont
  {Woo}}, \bibinfo {author} {\bibfnamefont {K.~M.}\ \bibnamefont {Song}},
  \bibinfo {author} {\bibfnamefont {H.-S.}\ \bibnamefont {Han}}, \bibinfo
  {author} {\bibfnamefont {M.-S.}\ \bibnamefont {Jung}}, \bibinfo {author}
  {\bibfnamefont {M.-Y.}\ \bibnamefont {Im}}, \bibinfo {author} {\bibfnamefont
  {K.-S.}\ \bibnamefont {Lee}}, \bibinfo {author} {\bibfnamefont {K.~S.}\
  \bibnamefont {Song}}, \bibinfo {author} {\bibfnamefont {P.}~\bibnamefont
  {Fischer}}, \bibinfo {author} {\bibfnamefont {J.-I.}\ \bibnamefont {Hong}},
  \bibinfo {author} {\bibfnamefont {J.~W.}\ \bibnamefont {Choi}}, \bibinfo
  {author} {\bibfnamefont {B.-C.}\ \bibnamefont {Min}}, \bibinfo {author}
  {\bibfnamefont {H.~C.}\ \bibnamefont {Koo}}, \ and\ \bibinfo {author}
  {\bibfnamefont {J.}~\bibnamefont {Chang}},\ }\href {\doibase
  10.1038/ncomms15573} {\bibfield  {journal} {\bibinfo  {journal} {Nat.
  Commun.}\ }\textbf {\bibinfo {volume} {8}},\ \bibinfo {pages} {15573}
  (\bibinfo {year} {2017})}\BibitemShut {NoStop}%
\bibitem [{\citenamefont {Jiang}\ \emph {et~al.}(2017)\citenamefont {Jiang},
  \citenamefont {Zhang}, \citenamefont {Yu}, \citenamefont {Zhang},
  \citenamefont {Wang}, \citenamefont {Jungfleisch}, \citenamefont {Pearson},
  \citenamefont {Cheng}, \citenamefont {Heinonen}, \citenamefont {Wang},
  \citenamefont {Zhou}, \citenamefont {Hoffmann},\ and\ \citenamefont
  {te~Velthuis}}]{Jiang2017}%
  \BibitemOpen
  \bibfield  {author} {\bibinfo {author} {\bibfnamefont {W.}~\bibnamefont
  {Jiang}}, \bibinfo {author} {\bibfnamefont {X.}~\bibnamefont {Zhang}},
  \bibinfo {author} {\bibfnamefont {G.}~\bibnamefont {Yu}}, \bibinfo {author}
  {\bibfnamefont {W.}~\bibnamefont {Zhang}}, \bibinfo {author} {\bibfnamefont
  {X.}~\bibnamefont {Wang}}, \bibinfo {author} {\bibfnamefont {M.~B.}\
  \bibnamefont {Jungfleisch}}, \bibinfo {author} {\bibfnamefont {J.~E.}\
  \bibnamefont {Pearson}}, \bibinfo {author} {\bibfnamefont {X.}~\bibnamefont
  {Cheng}}, \bibinfo {author} {\bibfnamefont {O.}~\bibnamefont {Heinonen}},
  \bibinfo {author} {\bibfnamefont {K.~L.}\ \bibnamefont {Wang}}, \bibinfo
  {author} {\bibfnamefont {Y.}~\bibnamefont {Zhou}}, \bibinfo {author}
  {\bibfnamefont {A.}~\bibnamefont {Hoffmann}}, \ and\ \bibinfo {author}
  {\bibfnamefont {S.~G.~E.}\ \bibnamefont {te~Velthuis}},\ }\href {\doibase
  10.1038/nphys3883} {\bibfield  {journal} {\bibinfo  {journal} {Nat. Phys.}\
  }\textbf {\bibinfo {volume} {13}},\ \bibinfo {pages} {162} (\bibinfo {year}
  {2017})}\BibitemShut {NoStop}%
\bibitem [{\citenamefont {Cowburn}\ \emph {et~al.}(1999)\citenamefont
  {Cowburn}, \citenamefont {Koltsov}, \citenamefont {Adeyeye},\ and\
  \citenamefont {Welland}}]{Cowburn1999}%
  \BibitemOpen
  \bibfield  {author} {\bibinfo {author} {\bibfnamefont {R.~P.}\ \bibnamefont
  {Cowburn}}, \bibinfo {author} {\bibfnamefont {D.~K.}\ \bibnamefont
  {Koltsov}}, \bibinfo {author} {\bibfnamefont {A.~O.}\ \bibnamefont
  {Adeyeye}}, \ and\ \bibinfo {author} {\bibfnamefont {M.~E.}\ \bibnamefont
  {Welland}},\ }\href {\doibase 10.1103/PhysRevLett.83.1042} {\bibfield
  {journal} {\bibinfo  {journal} {Phys. Rev. Lett.}\ }\textbf {\bibinfo
  {volume} {83}},\ \bibinfo {pages} {1042} (\bibinfo {year}
  {1999})}\BibitemShut {NoStop}%
\bibitem [{\citenamefont {Choe}\ \emph {et~al.}(2004)\citenamefont {Choe},
  \citenamefont {Acremann}, \citenamefont {Scholl}, \citenamefont {Bauer},
  \citenamefont {Doran}, \citenamefont {St{\"{o}}hr},\ and\ \citenamefont
  {Padmore}}]{Choe2004}%
  \BibitemOpen
  \bibfield  {author} {\bibinfo {author} {\bibfnamefont {S.~B.}\ \bibnamefont
  {Choe}}, \bibinfo {author} {\bibfnamefont {Y.}~\bibnamefont {Acremann}},
  \bibinfo {author} {\bibfnamefont {A.}~\bibnamefont {Scholl}}, \bibinfo
  {author} {\bibfnamefont {A.}~\bibnamefont {Bauer}}, \bibinfo {author}
  {\bibfnamefont {A.}~\bibnamefont {Doran}}, \bibinfo {author} {\bibfnamefont
  {J.}~\bibnamefont {St{\"{o}}hr}}, \ and\ \bibinfo {author} {\bibfnamefont
  {H.~A.}\ \bibnamefont {Padmore}},\ }\href {\doibase 10.1126/science.1095068}
  {\bibfield  {journal} {\bibinfo  {journal} {Science}\ }\textbf {\bibinfo
  {volume} {304}},\ \bibinfo {pages} {420} (\bibinfo {year}
  {2004})}\BibitemShut {NoStop}%
\bibitem [{\citenamefont {Buchanan}\ \emph {et~al.}(2005)\citenamefont
  {Buchanan}, \citenamefont {Roy}, \citenamefont {Grimsditch}, \citenamefont
  {Fradin}, \citenamefont {Guslienko}, \citenamefont {Bader},\ and\
  \citenamefont {Novosad}}]{Buchanan2005}%
  \BibitemOpen
  \bibfield  {author} {\bibinfo {author} {\bibfnamefont {K.~S.}\ \bibnamefont
  {Buchanan}}, \bibinfo {author} {\bibfnamefont {P.~E.}\ \bibnamefont {Roy}},
  \bibinfo {author} {\bibfnamefont {M.}~\bibnamefont {Grimsditch}}, \bibinfo
  {author} {\bibfnamefont {F.~Y.}\ \bibnamefont {Fradin}}, \bibinfo {author}
  {\bibfnamefont {K.~Y.}\ \bibnamefont {Guslienko}}, \bibinfo {author}
  {\bibfnamefont {S.~D.}\ \bibnamefont {Bader}}, \ and\ \bibinfo {author}
  {\bibfnamefont {V.}~\bibnamefont {Novosad}},\ }\href {\doibase
  10.1038/nphys173} {\bibfield  {journal} {\bibinfo  {journal} {Nat. Phys.}\
  }\textbf {\bibinfo {volume} {1}},\ \bibinfo {pages} {172} (\bibinfo {year}
  {2005})}\BibitemShut {NoStop}%
\bibitem [{\citenamefont {Kasai}\ \emph {et~al.}(2006)\citenamefont {Kasai},
  \citenamefont {Nakatani}, \citenamefont {Kobayashi}, \citenamefont {Kohno},\
  and\ \citenamefont {Ono}}]{Kasai2006}%
  \BibitemOpen
  \bibfield  {author} {\bibinfo {author} {\bibfnamefont {S.}~\bibnamefont
  {Kasai}}, \bibinfo {author} {\bibfnamefont {Y.}~\bibnamefont {Nakatani}},
  \bibinfo {author} {\bibfnamefont {K.}~\bibnamefont {Kobayashi}}, \bibinfo
  {author} {\bibfnamefont {H.}~\bibnamefont {Kohno}}, \ and\ \bibinfo {author}
  {\bibfnamefont {T.}~\bibnamefont {Ono}},\ }\href {\doibase
  10.1103/PhysRevLett.97.107204} {\bibfield  {journal} {\bibinfo  {journal}
  {Phys. Rev. Lett.}\ }\textbf {\bibinfo {volume} {97}},\ \bibinfo {pages}
  {107204} (\bibinfo {year} {2006})}\BibitemShut {NoStop}%
\bibitem [{\citenamefont {Caputo}\ \emph {et~al.}(2007)\citenamefont {Caputo},
  \citenamefont {Gaididei}, \citenamefont {Mertens},\ and\ \citenamefont
  {Sheka}}]{Caputo2007}%
  \BibitemOpen
  \bibfield  {author} {\bibinfo {author} {\bibfnamefont {J.-G.}\ \bibnamefont
  {Caputo}}, \bibinfo {author} {\bibfnamefont {Y.}~\bibnamefont {Gaididei}},
  \bibinfo {author} {\bibfnamefont {F.}~\bibnamefont {Mertens}}, \ and\
  \bibinfo {author} {\bibfnamefont {D.}~\bibnamefont {Sheka}},\ }\href
  {\doibase 10.1103/PhysRevLett.98.056604} {\bibfield  {journal} {\bibinfo
  {journal} {Phys. Rev. Lett.}\ }\textbf {\bibinfo {volume} {98}},\ \bibinfo
  {pages} {056604} (\bibinfo {year} {2007})}\BibitemShut {NoStop}%
\bibitem [{\citenamefont {Sugimoto}\ \emph {et~al.}(2011)\citenamefont
  {Sugimoto}, \citenamefont {Fukuma}, \citenamefont {Kasai}, \citenamefont
  {Kimura}, \citenamefont {Barman},\ and\ \citenamefont
  {Otani}}]{Sugimoto2011}%
  \BibitemOpen
  \bibfield  {author} {\bibinfo {author} {\bibfnamefont {S.}~\bibnamefont
  {Sugimoto}}, \bibinfo {author} {\bibfnamefont {Y.}~\bibnamefont {Fukuma}},
  \bibinfo {author} {\bibfnamefont {S.}~\bibnamefont {Kasai}}, \bibinfo
  {author} {\bibfnamefont {T.}~\bibnamefont {Kimura}}, \bibinfo {author}
  {\bibfnamefont {A.}~\bibnamefont {Barman}}, \ and\ \bibinfo {author}
  {\bibfnamefont {Y.}~\bibnamefont {Otani}},\ }\href {\doibase
  10.1103/PhysRevLett.106.197203} {\bibfield  {journal} {\bibinfo  {journal}
  {Phys. Rev. Lett.}\ }\textbf {\bibinfo {volume} {106}},\ \bibinfo {pages}
  {197203} (\bibinfo {year} {2011})}\BibitemShut {NoStop}%
\bibitem [{\citenamefont {Pollard}\ \emph {et~al.}(2012)\citenamefont
  {Pollard}, \citenamefont {Huang}, \citenamefont {Buchanan}, \citenamefont
  {Arena},\ and\ \citenamefont {Zhu}}]{Pollard2012}%
  \BibitemOpen
  \bibfield  {author} {\bibinfo {author} {\bibfnamefont {S.~D.}\ \bibnamefont
  {Pollard}}, \bibinfo {author} {\bibfnamefont {L.}~\bibnamefont {Huang}},
  \bibinfo {author} {\bibfnamefont {K.~S.}\ \bibnamefont {Buchanan}}, \bibinfo
  {author} {\bibfnamefont {D.~A.}\ \bibnamefont {Arena}}, \ and\ \bibinfo
  {author} {\bibfnamefont {Y.}~\bibnamefont {Zhu}},\ }\href {\doibase
  10.1038/ncomms2025} {\bibfield  {journal} {\bibinfo  {journal} {Nat.
  Commun.}\ }\textbf {\bibinfo {volume} {3}},\ \bibinfo {pages} {1028}
  (\bibinfo {year} {2012})}\BibitemShut {NoStop}%
\bibitem [{\citenamefont {{Van Waeyenberge}}\ \emph {et~al.}(2006)\citenamefont
  {{Van Waeyenberge}}, \citenamefont {Puzic}, \citenamefont {Stoll},
  \citenamefont {Chou}, \citenamefont {Tyliszczak}, \citenamefont {Hertel},
  \citenamefont {F{\"{a}}hnle}, \citenamefont {Br{\"{u}}ckl}, \citenamefont
  {Rott}, \citenamefont {Reiss}, \citenamefont {Neudecker}, \citenamefont
  {Weiss}, \citenamefont {Back},\ and\ \citenamefont
  {Sch{\"{u}}tz}}]{VanWaeyenberge2006}%
  \BibitemOpen
  \bibfield  {author} {\bibinfo {author} {\bibfnamefont {B.}~\bibnamefont {{Van
  Waeyenberge}}}, \bibinfo {author} {\bibfnamefont {A.}~\bibnamefont {Puzic}},
  \bibinfo {author} {\bibfnamefont {H.}~\bibnamefont {Stoll}}, \bibinfo
  {author} {\bibfnamefont {K.~W.}\ \bibnamefont {Chou}}, \bibinfo {author}
  {\bibfnamefont {T.}~\bibnamefont {Tyliszczak}}, \bibinfo {author}
  {\bibfnamefont {R.}~\bibnamefont {Hertel}}, \bibinfo {author} {\bibfnamefont
  {M.}~\bibnamefont {F{\"{a}}hnle}}, \bibinfo {author} {\bibfnamefont
  {H.}~\bibnamefont {Br{\"{u}}ckl}}, \bibinfo {author} {\bibfnamefont
  {K.}~\bibnamefont {Rott}}, \bibinfo {author} {\bibfnamefont {G.}~\bibnamefont
  {Reiss}}, \bibinfo {author} {\bibfnamefont {I.}~\bibnamefont {Neudecker}},
  \bibinfo {author} {\bibfnamefont {D.}~\bibnamefont {Weiss}}, \bibinfo
  {author} {\bibfnamefont {C.~H.}\ \bibnamefont {Back}}, \ and\ \bibinfo
  {author} {\bibfnamefont {G.}~\bibnamefont {Sch{\"{u}}tz}},\ }\href {\doibase
  10.1038/nature05240} {\bibfield  {journal} {\bibinfo  {journal} {Nature}\
  }\textbf {\bibinfo {volume} {444}},\ \bibinfo {pages} {461} (\bibinfo {year}
  {2006})}\BibitemShut {NoStop}%
\bibitem [{\citenamefont {Hasegawa}\ \emph {et~al.}(2017)\citenamefont
  {Hasegawa}, \citenamefont {Kondou}, \citenamefont {Kimata},\ and\
  \citenamefont {Otani}}]{Hasegawa2017}%
  \BibitemOpen
  \bibfield  {author} {\bibinfo {author} {\bibfnamefont {N.}~\bibnamefont
  {Hasegawa}}, \bibinfo {author} {\bibfnamefont {K.}~\bibnamefont {Kondou}},
  \bibinfo {author} {\bibfnamefont {M.}~\bibnamefont {Kimata}}, \ and\ \bibinfo
  {author} {\bibfnamefont {Y.}~\bibnamefont {Otani}},\ }\href {\doibase
  10.7567/APEX.10.053002} {\bibfield  {journal} {\bibinfo  {journal} {Appl.
  Phys. Express}\ }\textbf {\bibinfo {volume} {10}},\ \bibinfo {pages} {053002}
  (\bibinfo {year} {2017})}\BibitemShut {NoStop}%
\bibitem [{\citenamefont {Behncke}\ \emph {et~al.}(2018)\citenamefont
  {Behncke}, \citenamefont {Adolff}, \citenamefont {Wintz}, \citenamefont
  {H{\"{a}}nze}, \citenamefont {Schulte}, \citenamefont {Weigand},
  \citenamefont {Finizio}, \citenamefont {Raabe},\ and\ \citenamefont
  {Meier}}]{Behncke2018}%
  \BibitemOpen
  \bibfield  {author} {\bibinfo {author} {\bibfnamefont {C.}~\bibnamefont
  {Behncke}}, \bibinfo {author} {\bibfnamefont {C.~F.}\ \bibnamefont {Adolff}},
  \bibinfo {author} {\bibfnamefont {S.}~\bibnamefont {Wintz}}, \bibinfo
  {author} {\bibfnamefont {M.}~\bibnamefont {H{\"{a}}nze}}, \bibinfo {author}
  {\bibfnamefont {B.}~\bibnamefont {Schulte}}, \bibinfo {author} {\bibfnamefont
  {M.}~\bibnamefont {Weigand}}, \bibinfo {author} {\bibfnamefont
  {S.}~\bibnamefont {Finizio}}, \bibinfo {author} {\bibfnamefont
  {J.}~\bibnamefont {Raabe}}, \ and\ \bibinfo {author} {\bibfnamefont
  {G.}~\bibnamefont {Meier}},\ }\href {\doibase 10.1038/s41598-017-17480-1}
  {\bibfield  {journal} {\bibinfo  {journal} {Sci. Rep.}\ }\textbf {\bibinfo
  {volume} {8}},\ \bibinfo {pages} {186} (\bibinfo {year} {2018})}\BibitemShut
  {NoStop}%
\bibitem [{\citenamefont {Mihajlovi{\'{c}}}\ \emph {et~al.}(2010)\citenamefont
  {Mihajlovi{\'{c}}}, \citenamefont {Patrick}, \citenamefont {Pearson},
  \citenamefont {Novosad}, \citenamefont {Bader}, \citenamefont {Field},
  \citenamefont {Sullivan},\ and\ \citenamefont {Hoffmann}}]{Mihajlovic2010}%
  \BibitemOpen
  \bibfield  {author} {\bibinfo {author} {\bibfnamefont {G.}~\bibnamefont
  {Mihajlovi{\'{c}}}}, \bibinfo {author} {\bibfnamefont {M.~S.}\ \bibnamefont
  {Patrick}}, \bibinfo {author} {\bibfnamefont {J.~E.}\ \bibnamefont
  {Pearson}}, \bibinfo {author} {\bibfnamefont {V.}~\bibnamefont {Novosad}},
  \bibinfo {author} {\bibfnamefont {S.~D.}\ \bibnamefont {Bader}}, \bibinfo
  {author} {\bibfnamefont {M.}~\bibnamefont {Field}}, \bibinfo {author}
  {\bibfnamefont {G.~J.}\ \bibnamefont {Sullivan}}, \ and\ \bibinfo {author}
  {\bibfnamefont {A.}~\bibnamefont {Hoffmann}},\ }\href {\doibase
  10.1063/1.3360841} {\bibfield  {journal} {\bibinfo  {journal} {Appl. Phys.
  Lett.}\ }\textbf {\bibinfo {volume} {96}},\ \bibinfo {pages} {112501}
  (\bibinfo {year} {2010})},\ \Eprint {http://arxiv.org/abs/0911.2267}
  {arXiv:0911.2267} \BibitemShut {NoStop}%
\bibitem [{\citenamefont {Mayadas}, \citenamefont {Janak},\ and\ \citenamefont
  {Gangulee}(1974)}]{Mayadas1974}%
  \BibitemOpen
  \bibfield  {author} {\bibinfo {author} {\bibfnamefont {A.~F.}\ \bibnamefont
  {Mayadas}}, \bibinfo {author} {\bibfnamefont {J.~F.}\ \bibnamefont {Janak}},
  \ and\ \bibinfo {author} {\bibfnamefont {A.}~\bibnamefont {Gangulee}},\
  }\href {\doibase 10.1063/1.1663668} {\bibfield  {journal} {\bibinfo
  {journal} {J. Appl. Phys.}\ }\textbf {\bibinfo {volume} {45}},\ \bibinfo
  {pages} {2780} (\bibinfo {year} {1974})}\BibitemShut {NoStop}%
\bibitem [{\citenamefont {Sushruth}\ \emph {et~al.}(2016)\citenamefont
  {Sushruth}, \citenamefont {Fried}, \citenamefont {Anane}, \citenamefont
  {Xavier}, \citenamefont {Deranlot}, \citenamefont {Kostylev}, \citenamefont
  {Cros},\ and\ \citenamefont {Metaxas}}]{Sushruth2016}%
  \BibitemOpen
  \bibfield  {author} {\bibinfo {author} {\bibfnamefont {M.}~\bibnamefont
  {Sushruth}}, \bibinfo {author} {\bibfnamefont {J.~P.}\ \bibnamefont {Fried}},
  \bibinfo {author} {\bibfnamefont {A.}~\bibnamefont {Anane}}, \bibinfo
  {author} {\bibfnamefont {S.}~\bibnamefont {Xavier}}, \bibinfo {author}
  {\bibfnamefont {C.}~\bibnamefont {Deranlot}}, \bibinfo {author}
  {\bibfnamefont {M.}~\bibnamefont {Kostylev}}, \bibinfo {author}
  {\bibfnamefont {V.}~\bibnamefont {Cros}}, \ and\ \bibinfo {author}
  {\bibfnamefont {P.~J.}\ \bibnamefont {Metaxas}},\ }\href {\doibase
  10.1103/PhysRevB.94.100402} {\bibfield  {journal} {\bibinfo  {journal} {Phys.
  Rev. B}\ }\textbf {\bibinfo {volume} {94}},\ \bibinfo {pages} {100402}
  (\bibinfo {year} {2016})},\ \Eprint {http://arxiv.org/abs/1605.01826}
  {arXiv:1605.01826} \BibitemShut {NoStop}%
\bibitem [{\citenamefont {Sushruth}\ \emph {et~al.}(2017)\citenamefont
  {Sushruth}, \citenamefont {Fried}, \citenamefont {Anane}, \citenamefont
  {Xavier}, \citenamefont {Deranlot}, \citenamefont {Cros},\ and\ \citenamefont
  {Metaxas}}]{Sushruth2017}%
  \BibitemOpen
  \bibfield  {author} {\bibinfo {author} {\bibfnamefont {M.}~\bibnamefont
  {Sushruth}}, \bibinfo {author} {\bibfnamefont {J.~P.}\ \bibnamefont {Fried}},
  \bibinfo {author} {\bibfnamefont {A.}~\bibnamefont {Anane}}, \bibinfo
  {author} {\bibfnamefont {S.}~\bibnamefont {Xavier}}, \bibinfo {author}
  {\bibfnamefont {C.}~\bibnamefont {Deranlot}}, \bibinfo {author}
  {\bibfnamefont {V.}~\bibnamefont {Cros}}, \ and\ \bibinfo {author}
  {\bibfnamefont {P.~J.}\ \bibnamefont {Metaxas}},\ }\href {\doibase
  10.1103/PhysRevB.96.060405} {\bibfield  {journal} {\bibinfo  {journal} {Phys.
  Rev. B}\ }\textbf {\bibinfo {volume} {96}},\ \bibinfo {pages} {060405}
  (\bibinfo {year} {2017})}\BibitemShut {NoStop}%
\bibitem [{\citenamefont {Cui}\ \emph {et~al.}(2015)\citenamefont {Cui},
  \citenamefont {Hu}, \citenamefont {Hidegara}, \citenamefont {Yakata},\ and\
  \citenamefont {Kimura}}]{Cui2015}%
  \BibitemOpen
  \bibfield  {author} {\bibinfo {author} {\bibfnamefont {X.}~\bibnamefont
  {Cui}}, \bibinfo {author} {\bibfnamefont {S.}~\bibnamefont {Hu}}, \bibinfo
  {author} {\bibfnamefont {M.}~\bibnamefont {Hidegara}}, \bibinfo {author}
  {\bibfnamefont {S.}~\bibnamefont {Yakata}}, \ and\ \bibinfo {author}
  {\bibfnamefont {T.}~\bibnamefont {Kimura}},\ }\href {\doibase
  10.1038/srep17922} {\bibfield  {journal} {\bibinfo  {journal} {Sci. Rep.}\
  }\textbf {\bibinfo {volume} {5}},\ \bibinfo {pages} {17922} (\bibinfo {year}
  {2015})}\BibitemShut {NoStop}%
\bibitem [{\citenamefont {{\v{S}}{\v{c}}epka}\ \emph
  {et~al.}(2015)\citenamefont {{\v{S}}{\v{c}}epka}, \citenamefont
  {Polakovi{\v{c}}}, \citenamefont {{\v{S}}olt{\'{y}}s}, \citenamefont
  {T{\'{o}}bik}, \citenamefont {Kulich}, \citenamefont {K{\'{u}}dela},
  \citenamefont {D{\'{e}}rer},\ and\ \citenamefont {Cambel}}]{Scepka2015}%
  \BibitemOpen
  \bibfield  {author} {\bibinfo {author} {\bibfnamefont {T.}~\bibnamefont
  {{\v{S}}{\v{c}}epka}}, \bibinfo {author} {\bibfnamefont {T.}~\bibnamefont
  {Polakovi{\v{c}}}}, \bibinfo {author} {\bibfnamefont {J.}~\bibnamefont
  {{\v{S}}olt{\'{y}}s}}, \bibinfo {author} {\bibfnamefont {J.}~\bibnamefont
  {T{\'{o}}bik}}, \bibinfo {author} {\bibfnamefont {M.}~\bibnamefont {Kulich}},
  \bibinfo {author} {\bibfnamefont {R.}~\bibnamefont {K{\'{u}}dela}}, \bibinfo
  {author} {\bibfnamefont {J.}~\bibnamefont {D{\'{e}}rer}}, \ and\ \bibinfo
  {author} {\bibfnamefont {V.}~\bibnamefont {Cambel}},\ }\href {\doibase
  10.1063/1.4935437} {\bibfield  {journal} {\bibinfo  {journal} {AIP Adv.}\
  }\textbf {\bibinfo {volume} {5}},\ \bibinfo {pages} {117205} (\bibinfo {year}
  {2015})}\BibitemShut {NoStop}%
\bibitem [{Note1()}]{Note1}%
  \BibitemOpen
  \bibinfo {note} {M.R. Scheinfein, http://llgmicro.home.mindspring.com
  (1997).}\BibitemShut {Stop}%
\bibitem [{\citenamefont {Im}\ \emph {et~al.}(2012)\citenamefont {Im},
  \citenamefont {Fischer}, \citenamefont {Yamada}, \citenamefont {Sato},
  \citenamefont {Kasai}, \citenamefont {Nakatani},\ and\ \citenamefont
  {Ono}}]{Im2012}%
  \BibitemOpen
  \bibfield  {author} {\bibinfo {author} {\bibfnamefont {M.-Y.}\ \bibnamefont
  {Im}}, \bibinfo {author} {\bibfnamefont {P.}~\bibnamefont {Fischer}},
  \bibinfo {author} {\bibfnamefont {K.}~\bibnamefont {Yamada}}, \bibinfo
  {author} {\bibfnamefont {T.}~\bibnamefont {Sato}}, \bibinfo {author}
  {\bibfnamefont {S.}~\bibnamefont {Kasai}}, \bibinfo {author} {\bibfnamefont
  {Y.}~\bibnamefont {Nakatani}}, \ and\ \bibinfo {author} {\bibfnamefont
  {T.}~\bibnamefont {Ono}},\ }\href {\doibase 10.1038/ncomms1978} {\bibfield
  {journal} {\bibinfo  {journal} {Nat. Commun.}\ }\textbf {\bibinfo {volume}
  {3}},\ \bibinfo {pages} {983} (\bibinfo {year} {2012})}\BibitemShut {NoStop}%
\bibitem [{\citenamefont {Luo}\ \emph {et~al.}(2014)\citenamefont {Luo},
  \citenamefont {Zhou}, \citenamefont {Won},\ and\ \citenamefont
  {Wu}}]{Luo2014}%
  \BibitemOpen
  \bibfield  {author} {\bibinfo {author} {\bibfnamefont {Y.~M.}\ \bibnamefont
  {Luo}}, \bibinfo {author} {\bibfnamefont {C.}~\bibnamefont {Zhou}}, \bibinfo
  {author} {\bibfnamefont {C.}~\bibnamefont {Won}}, \ and\ \bibinfo {author}
  {\bibfnamefont {Y.~Z.}\ \bibnamefont {Wu}},\ }\href {\doibase
  10.1063/1.4874135} {\bibfield  {journal} {\bibinfo  {journal} {AIP Adv.}\
  }\textbf {\bibinfo {volume} {4}},\ \bibinfo {pages} {047136} (\bibinfo {year}
  {2014})}\BibitemShut {NoStop}%
\bibitem [{\citenamefont {Luo}\ \emph {et~al.}(2015)\citenamefont {Luo},
  \citenamefont {Zhou}, \citenamefont {Won},\ and\ \citenamefont
  {Wu}}]{Luo2015}%
  \BibitemOpen
  \bibfield  {author} {\bibinfo {author} {\bibfnamefont {Y.~M.}\ \bibnamefont
  {Luo}}, \bibinfo {author} {\bibfnamefont {C.}~\bibnamefont {Zhou}}, \bibinfo
  {author} {\bibfnamefont {C.}~\bibnamefont {Won}}, \ and\ \bibinfo {author}
  {\bibfnamefont {Y.~Z.}\ \bibnamefont {Wu}},\ }\href {\doibase
  10.1063/1.4919423} {\bibfield  {journal} {\bibinfo  {journal} {J. Appl.
  Phys.}\ }\textbf {\bibinfo {volume} {117}},\ \bibinfo {pages} {163916}
  (\bibinfo {year} {2015})}\BibitemShut {NoStop}%
\bibitem [{\citenamefont {Guslienko}\ \emph {et~al.}(2006)\citenamefont
  {Guslienko}, \citenamefont {Han}, \citenamefont {Keavney}, \citenamefont
  {Divan},\ and\ \citenamefont {Bader}}]{Guslienko2006}%
  \BibitemOpen
  \bibfield  {author} {\bibinfo {author} {\bibfnamefont {K.~Y.}\ \bibnamefont
  {Guslienko}}, \bibinfo {author} {\bibfnamefont {X.~F.}\ \bibnamefont {Han}},
  \bibinfo {author} {\bibfnamefont {D.~J.}\ \bibnamefont {Keavney}}, \bibinfo
  {author} {\bibfnamefont {R.}~\bibnamefont {Divan}}, \ and\ \bibinfo {author}
  {\bibfnamefont {S.~D.}\ \bibnamefont {Bader}},\ }\href {\doibase
  10.1103/PhysRevLett.96.067205} {\bibfield  {journal} {\bibinfo  {journal}
  {Phys. Rev. Lett.}\ }\textbf {\bibinfo {volume} {96}},\ \bibinfo {pages}
  {067205} (\bibinfo {year} {2006})}\BibitemShut {NoStop}%
\bibitem [{\citenamefont {Burgess}\ \emph {et~al.}(2013)\citenamefont
  {Burgess}, \citenamefont {Fraser}, \citenamefont {Sani}, \citenamefont
  {Vick}, \citenamefont {Hauer}, \citenamefont {Davis},\ and\ \citenamefont
  {Freeman}}]{Burgess2013}%
  \BibitemOpen
  \bibfield  {author} {\bibinfo {author} {\bibfnamefont {J.~A.~J.}\
  \bibnamefont {Burgess}}, \bibinfo {author} {\bibfnamefont {A.~E.}\
  \bibnamefont {Fraser}}, \bibinfo {author} {\bibfnamefont {F.~F.}\
  \bibnamefont {Sani}}, \bibinfo {author} {\bibfnamefont {D.}~\bibnamefont
  {Vick}}, \bibinfo {author} {\bibfnamefont {B.~D.}\ \bibnamefont {Hauer}},
  \bibinfo {author} {\bibfnamefont {J.~P.}\ \bibnamefont {Davis}}, \ and\
  \bibinfo {author} {\bibfnamefont {M.~R.}\ \bibnamefont {Freeman}},\ }\href
  {\doibase 10.1126/science.1231390} {\bibfield  {journal} {\bibinfo  {journal}
  {Science}\ }\textbf {\bibinfo {volume} {339}},\ \bibinfo {pages} {1051 LP }
  (\bibinfo {year} {2013})}\BibitemShut {NoStop}%
\bibitem [{\citenamefont {Jenkins}\ \emph {et~al.}(2019)\citenamefont
  {Jenkins}, \citenamefont {Alvarez}, \citenamefont {Freitas},\ and\
  \citenamefont {Ferreira}}]{Jenkins2019}%
  \BibitemOpen
  \bibfield  {author} {\bibinfo {author} {\bibfnamefont {A.~S.}\ \bibnamefont
  {Jenkins}}, \bibinfo {author} {\bibfnamefont {L.~S.~E.}\ \bibnamefont
  {Alvarez}}, \bibinfo {author} {\bibfnamefont {P.~P.}\ \bibnamefont
  {Freitas}}, \ and\ \bibinfo {author} {\bibfnamefont {R.}~\bibnamefont
  {Ferreira}},\ }\href {\doibase 10.1038/s41598-019-52236-z} {\bibfield
  {journal} {\bibinfo  {journal} {Sci. Rep.}\ }\textbf {\bibinfo {volume}
  {9}},\ \bibinfo {pages} {15661} (\bibinfo {year} {2019})}\BibitemShut
  {NoStop}%
\bibitem [{\citenamefont {Lu}\ \emph {et~al.}(2013)\citenamefont {Lu},
  \citenamefont {Wu}, \citenamefont {Mallary}, \citenamefont {Bertero},
  \citenamefont {Srinivasan}, \citenamefont {Acharya}, \citenamefont
  {Schulthei{\ss}},\ and\ \citenamefont {Hoffmann}}]{Lu2013}%
  \BibitemOpen
  \bibfield  {author} {\bibinfo {author} {\bibfnamefont {L.}~\bibnamefont
  {Lu}}, \bibinfo {author} {\bibfnamefont {M.}~\bibnamefont {Wu}}, \bibinfo
  {author} {\bibfnamefont {M.}~\bibnamefont {Mallary}}, \bibinfo {author}
  {\bibfnamefont {G.}~\bibnamefont {Bertero}}, \bibinfo {author} {\bibfnamefont
  {K.}~\bibnamefont {Srinivasan}}, \bibinfo {author} {\bibfnamefont
  {R.}~\bibnamefont {Acharya}}, \bibinfo {author} {\bibfnamefont
  {H.}~\bibnamefont {Schulthei{\ss}}}, \ and\ \bibinfo {author} {\bibfnamefont
  {A.}~\bibnamefont {Hoffmann}},\ }\href {\doibase 10.1063/1.4816798}
  {\bibfield  {journal} {\bibinfo  {journal} {Appl. Phys. Lett.}\ }\textbf
  {\bibinfo {volume} {103}},\ \bibinfo {pages} {042413} (\bibinfo {year}
  {2013})}\BibitemShut {NoStop}%
\bibitem [{\citenamefont {Shinjo}\ \emph {et~al.}(2000)\citenamefont {Shinjo},
  \citenamefont {Okuno}, \citenamefont {Hassdorf}, \citenamefont {Shigeto},\
  and\ \citenamefont {Ono}}]{Shinjo2000}%
  \BibitemOpen
  \bibfield  {author} {\bibinfo {author} {\bibfnamefont {T.}~\bibnamefont
  {Shinjo}}, \bibinfo {author} {\bibfnamefont {T.}~\bibnamefont {Okuno}},
  \bibinfo {author} {\bibfnamefont {R.}~\bibnamefont {Hassdorf}}, \bibinfo
  {author} {\bibfnamefont {K.}\ \bibnamefont {Shigeto}}, \ and\ \bibinfo
  {author} {\bibfnamefont {T.}~\bibnamefont {Ono}},\ }\href {\doibase
  10.1126/science.289.5481.930} {\bibfield  {journal} {\bibinfo  {journal}
  {Science}\ }\textbf {\bibinfo {volume} {289}},\ \bibinfo {pages} {930}
  (\bibinfo {year} {2000})}\BibitemShut {NoStop}%
\bibitem [{\citenamefont {Bisig}\ \emph {et~al.}(2016)\citenamefont {Bisig},
  \citenamefont {Akosa}, \citenamefont {Moon}, \citenamefont {Rhensius},
  \citenamefont {Moutafis}, \citenamefont {von Bieren}, \citenamefont
  {Heidler}, \citenamefont {Kiliani}, \citenamefont {Kammerer}, \citenamefont
  {Curcic}, \citenamefont {Weigand}, \citenamefont {Tyliszczak}, \citenamefont
  {{Van Waeyenberge}}, \citenamefont {Stoll}, \citenamefont {Sch{\"{u}}tz},
  \citenamefont {Lee}, \citenamefont {Manchon},\ and\ \citenamefont
  {Kl{\"{a}}ui}}]{Bisig2016}%
  \BibitemOpen
  \bibfield  {author} {\bibinfo {author} {\bibfnamefont {A.}~\bibnamefont
  {Bisig}}, \bibinfo {author} {\bibfnamefont {C.~A.}\ \bibnamefont {Akosa}},
  \bibinfo {author} {\bibfnamefont {J.-H.}\ \bibnamefont {Moon}}, \bibinfo
  {author} {\bibfnamefont {J.}~\bibnamefont {Rhensius}}, \bibinfo {author}
  {\bibfnamefont {C.}~\bibnamefont {Moutafis}}, \bibinfo {author}
  {\bibfnamefont {A.}~\bibnamefont {von Bieren}}, \bibinfo {author}
  {\bibfnamefont {J.}~\bibnamefont {Heidler}}, \bibinfo {author} {\bibfnamefont
  {G.}~\bibnamefont {Kiliani}}, \bibinfo {author} {\bibfnamefont
  {M.}~\bibnamefont {Kammerer}}, \bibinfo {author} {\bibfnamefont
  {M.}~\bibnamefont {Curcic}}, \bibinfo {author} {\bibfnamefont
  {M.}~\bibnamefont {Weigand}}, \bibinfo {author} {\bibfnamefont
  {T.}~\bibnamefont {Tyliszczak}}, \bibinfo {author} {\bibfnamefont
  {B.}~\bibnamefont {{Van Waeyenberge}}}, \bibinfo {author} {\bibfnamefont
  {H.}~\bibnamefont {Stoll}}, \bibinfo {author} {\bibfnamefont
  {G.}~\bibnamefont {Sch{\"{u}}tz}}, \bibinfo {author} {\bibfnamefont {K.-J.}\
  \bibnamefont {Lee}}, \bibinfo {author} {\bibfnamefont {A.}~\bibnamefont
  {Manchon}}, \ and\ \bibinfo {author} {\bibfnamefont {M.}~\bibnamefont
  {Kl{\"{a}}ui}},\ }\href {\doibase 10.1103/PhysRevLett.117.277203} {\bibfield
  {journal} {\bibinfo  {journal} {Phys. Rev. Lett.}\ }\textbf {\bibinfo
  {volume} {117}},\ \bibinfo {pages} {277203} (\bibinfo {year}
  {2016})}\BibitemShut {NoStop}%
\bibitem [{\citenamefont {Haidar}\ \emph {et~al.}(2019)\citenamefont {Haidar},
  \citenamefont {Awad}, \citenamefont {Dvornik}, \citenamefont {Khymyn},
  \citenamefont {Houshang},\ and\ \citenamefont {{\AA}kerman}}]{Haidar2019}%
  \BibitemOpen
  \bibfield  {author} {\bibinfo {author} {\bibfnamefont {M.}~\bibnamefont
  {Haidar}}, \bibinfo {author} {\bibfnamefont {A.~A.}\ \bibnamefont {Awad}},
  \bibinfo {author} {\bibfnamefont {M.}~\bibnamefont {Dvornik}}, \bibinfo
  {author} {\bibfnamefont {R.}~\bibnamefont {Khymyn}}, \bibinfo {author}
  {\bibfnamefont {A.}~\bibnamefont {Houshang}}, \ and\ \bibinfo {author}
  {\bibfnamefont {J.}~\bibnamefont {{\AA}kerman}},\ }\href {\doibase
  10.1038/s41467-019-10120-4} {\bibfield  {journal} {\bibinfo  {journal} {Nat.
  Commun.}\ }\textbf {\bibinfo {volume} {10}},\ \bibinfo {pages} {2362}
  (\bibinfo {year} {2019})}\BibitemShut {NoStop}%
\end{thebibliography}
\end{document}